\documentclass[twocolumn,american,nofootinbib]{revtex4}
\usepackage{ae}
\usepackage{aecompl}
\usepackage[T1]{fontenc}
\usepackage[latin1]{inputenc}
\usepackage{amsmath}
\usepackage{amssymb}

\makeatletter
\usepackage{psfrag}

\usepackage{babel}
\makeatother
\begin{document}

\title{Attractor scenarios and superluminal signals in $k$-essence cosmology}

\author{Jin U Kang$^{1,2}$, Vitaly Vanchurin$^{2}$, and Sergei Winitzki$^{2}$}

\affiliation{$^{1}$Department of Physics, Kim Il Sung University, Pyongyang, DPR Korea}

\affiliation{$^{2}$Arnold Sommerfeld Center, Department of Physics, Ludwig-Maximilians
University, Theresienstr.~37, 80333 Munich, Germany}

\begin{abstract}
Cosmological scenarios with $k$-essence are invoked in order to explain the
observed late-time acceleration of the universe. These scenarios avoid the need
for fine-tuned initial conditions (the {}``coincidence problem'') because
of the attractor-like dynamics of the $k$-essence field $\phi$. It was recently
shown that all $k$-essence scenarios with Lagrangians $p=L(X)\phi^{-2}$, where
$X\equiv\frac{1}{2}\phi_{,\mu}\phi^{,\mu}$, necessarily involve an epoch where
perturbations of $\phi$ propagate faster than light (the {}``no-go theorem'').
We carry out a comprehensive study of attractor-like cosmological solutions
({}``trackers'') involving a $k$-essence scalar field $\phi$ and another
matter component. The result of this study is a complete classification of $k$-essence
Lagrangians that admit asymptotically stable tracking solutions, among all Lagrangians
of the form $p=K(\phi)L(X)$. Using this classification, we select the class
of models that describe the late-time acceleration and avoid the coincidence
problem through the tracking mechanism. An analogous {}``no-go theorem'' still
holds for this class of models, indicating the existence of a superluminal epoch.
In the context of $k$-essence cosmology, the superluminal epoch does not lead
to causality violations. We discuss the implications of superluminal signal
propagation for possible causality violations in Lorentz-invariant field theories.
\end{abstract}
\maketitle

\section{Introduction and overview of results}

Cosmological scenarios involving a scalar field known as $k$-essence~\cite{Chiba:1999ka,Armendariz-Picon:2000dh,Armendariz-Picon:2000ah}
are intended to explain the late-time acceleration of the universe (see Ref.~\cite{Copeland:2006wr}
for a recent review of dynamical models of dark energy). An important motivation
behind the $k$-essence scenarios is to avoid the fine-tuning of the initial
conditions for the scalar field (the {}``coincidence problem'').

The effective Lagrangian $p(X,\phi)$ describing the dynamics of the scalar
field $\phi$ consists of a noncanonical kinetic term, \begin{equation}
p(X,\phi)=K(\phi)L(X),\quad X\equiv\frac{1}{2}\partial_{\mu}\phi\partial^{\mu}\phi,\label{eq:Kessence L}\end{equation}
where $K(\phi)$ and $L(X)$ are functions determined by the underlying fundamental
theory. One considers the evolution of the field $\phi$ coupled to gravity
in a standard homogeneous cosmology in the presence of matter. With a suitable
choice of the Lagrangian, the evolution of $\phi$ during radiation domination
quickly drives the system into a region in phase space where the $k$-essence
field $\phi$ has a nearly constant equation of state with $w_{\phi}=\frac{1}{3}$,
mimicking radiation. Thus the energy density $\varepsilon_{\phi}$ of $k$-essence
approaches a constant fraction of the energy density $\varepsilon_{m}$ of the
radiation. This behavior of $k$-essence ($w_{\phi}\rightarrow\textrm{const}$
and $\varepsilon_{\phi}/\varepsilon_{\text{tot}}\rightarrow\textrm{const}$,
where $\varepsilon_{\text{tot}}\equiv\varepsilon_{\phi}+\varepsilon_{m}$) is
called tracking, and the solution with $w_{\phi}\approx\textrm{const}$ is called
a tracker solution. 

The parameters of the Lagrangian can be adjusted such that the energy density
in $k$-essence during the radiation era is small ($\varepsilon_{m}\approx\varepsilon_{\text{tot}}$),
so that the standard cosmological evolution is not significantly altered. After
the onset of dust domination ($w_{m}=0$), the energy density in $k$-essence
quickly becomes negligible and the evolution leaves the radiation tracker. A
tracking solution with $w_{\phi}=0$ does not exist (due to a particular choice
of the Lagrangian), and instead the $k$-essence is driven to a tracking regime
with $w_{\phi}\approx\textrm{const}<0$. Since $w_{\phi}<w_{m}$, the $k$-essence
will eventually dominate the energy density of the dust component. The precise
value of $w_{\phi}$ in that regime can be parametrically adjusted to fit the
currently observed data; in particular, values $w_{\phi}\approx-1$ can be achieved.%
\footnote{We note that the {}``phantom'' values $w_{\phi}<-1$ cannot be reached in
this single-field model; see e.g.~\cite{Vikman:2004dc,Caldwell:2005ai}. {}``Phantom''
models such as that of Ref.~\cite{Hao:2003ib} cannot describe the tracking
behavior of $k$-essence since in these models $w_{\phi}<-1$ at all times.%
}

In our terminology, a {}``tracking solution'' is a solution for which $w_{\phi}$
approaches a fixed value, whether or not this value is equal to the equation
of state parameter $w_{m}$ of the dominant matter component. It is essential
that the tracker solutions are stable attractors for nearby solutions. Because
of this property, the field $\phi$ is driven into the tracker regime in the
phase space with fixed values of $w_{\phi}$ and the ratio $\varepsilon_{\phi}/\varepsilon_{\text{tot}}$,
for a wide range of initial conditions for $\phi$. To construct a viable $k$-essence
model, it is important to choose a Lagrangian $p(X,\phi)$ for which stable
tracker solutions exist within the radiation- and dust-dominated cosmological
eras.

Previous works concerning the dynamics of $k$-essence either assumed a specific
form of the Lagrangian, for instance~\cite{Armendariz-Picon:2000ah}\begin{equation}
p(X,\phi)=\frac{L(X)}{\phi^{2}},\label{eq:quadratic phi}\end{equation}
or imposed ad hoc restrictions on the Lagrangian with the purpose of deriving
exact solutions (e.g.~\cite{Bertacca:2007ux}). In particular, it was assumed
that $w_{\phi}=\textrm{const}$ is an exact solution of the equations of motion.
However, the physically necessary requirement is weaker: namely, one merely
needs that $w_{\phi}$ should \emph{approach} a constant value asymptotically
at late times. The existence of an \emph{exact} solution $w_{\phi}=\textrm{const}$
is not necessary. With this weaker requirement, a much wider range of Lagrangians
enters the consideration. 

In the present paper, we restrict our attention to Lagrangians of the {}``factorized''
form~(\ref{eq:Kessence L}) but do not impose any further a priori restrictions
on the Lagrangians; neither do we require the existence of analytic exact solutions,
or of solutions with $w_{\phi}=\textrm{const}$. It is only assumed that the
cosmological scenario is realized with $\dot{\phi}>0$ and that $\phi$ reaches
arbitrarily large values. Our results can be viewed as a comprehensive extension
of previous studies of attractor behavior in $k$-essence cosmology (e.g.~\cite{Chiba:2002mw,Das:2006cm,Bertacca:2007ux}).
We determine the class of Lagrangians $p(X,\phi)$ that admit stable tracking
regimes in which $w_{\phi}\rightarrow\textrm{const}$, for a given value of
$w_{m}$. The possible asymptotic values of $w_{\phi}$ and $\varepsilon_{\phi}/\varepsilon_{\text{tot}}$
are derived in each case. 

The form~(\ref{eq:Kessence L}) is sufficiently general to reproduce an observationally
measured cosmological history~\cite{Li:2006bx} and covers many interesting
cases, such as $k$-essence with purely kinetic term~\cite{Scherrer:2004au}
or the {}``kinetic quintessence''~\cite{Chiba:1999ka}. Factorized Lagrangians
have been the main focus of attention in the study of $k$-essence (see e.g.~\cite{Malquarti:2003nn,Wei:2004rw,Rendall:2005fv}).
More generally, Lagrangians of the form\begin{equation}
p(X,\phi)=\left[K_{1}(\phi)X^{n_{1}}+K_{2}(\phi)X^{n_{2}}\right]^{n_{3}},\end{equation}
where $n_{1},n_{2},n_{3}$ are constants, can be reduced to the Lagrangian~(\ref{eq:Kessence L})
by a suitable redefinition of the field $\phi$. Our analysis will also apply
to Lagrangians that have the asymptotic form $p\approx K(\phi)L(X)$ for $\phi\rightarrow\infty$
and for which only the large-$\phi$ regime is cosmologically relevant. Nonfactorizable
Lagrangians, such as those studied in Refs.~\cite{Silverstein:2003hf,Alishahiha:2004eh,Calcagni:2006ge,Fang:2006yh,Bertacca:2007ux},
require a separate consideration which we do not attempt here.

Recently, it was shown that the scenarios of $k$-essence cosmology with Lagrangians
of the form~(\ref{eq:quadratic phi}) necessarily include an epoch when perturbations
in the $k$-essence field propagate faster than light (the {}``no-go theorem''~\cite{Bonvin:2006vc}).
It is well known that superluminal propagation of perturbations opens the possibility
of causality violations, although causality is actually preserved in many cases.
This issue has been a subject of some debate, see e.g.~the discussion in Refs.~\cite{Liberati:2001sd,Adams:2006sv,Babichev:2006vx,Dubovsky:2006vk,Bruneton:2006gf,Babichev:2007wg,Ellis:2007ic,Bonvin:2007mw}.
One of the motivations for the present work is to determine whether the {}``no-go
theorem,'' derived for a restricted class of $k$-essence Lagrangians, still
holds in scenarios with more general Lagrangians.

To answer this question, we performed an exhaustive analysis of all the possibilities
for the existence of stable tracking solutions in ghost-free $k$-essence theories
with positive energy density (the complete list of physical restrictions is
given in Sec.~\ref{sec:Physical restrictions}). We considered the cosmological
evolution of a scalar $k$-essence field $\phi$ coupled through gravity to
a matter component having a fixed equation-of state parameter $w_{m}$. In this
context, we enumerated all Lagrangians of the form~(\ref{eq:Kessence L}) that
admit attractor solutions with $w_{\phi}\rightarrow\textrm{const}$ and $\varepsilon_{\phi}/\varepsilon_{\text{tot}}\rightarrow\textrm{const}$
at late times (Sec.~\ref{sec:Viable-Lagrangians-for}). Since our task is to
determine the entire class of theories admitting a certain asymptotic behavior,
numerical calculations could not be used. The analytic method used for the asymptotic
analysis of the dynamical evolution is outlined at the beginning of Appendix~\ref{sec:Asymptotically-stable-solutions},
where all the calculations are presented in detail. This method is similar to
that developed in Ref.~\cite{Helmer:2006tz} for the analysis of attractors
in models of $k$-inflation.

Armed with the complete enumeration of stable trackers, we then selected the
Lagrangians capable of providing a subdominant tracker solution during the radiation
era and an asymptotically dominant tracker solution during the dust era. We
show that the only appropriate class of Lagrangians consists of functions $p(X,\phi)$
of the form\begin{equation}
p(X,\phi)=\frac{1+K_{0}(\phi)}{\phi^{2}}L(X),\quad\lim_{\phi\rightarrow\infty}K_{0}(\phi)=0.\end{equation}
Since the dynamical evolution drives $\phi$ towards very large values, these
Lagrangians are practically indistinguishable from the Lagrangians of the form~(\ref{eq:quadratic phi}).
Then one can prove, similarly to Ref.~\cite{Bonvin:2006vc}, that the cosmological
evolution necessarily includes an epoch where perturbations of the $k$-essence
field $\phi$ propagate with a superluminal speed. Thus, we prove the {}``no-go
theorem'' starting from a much wider initial class of $k$-essence Lagrangians.

In Sec.~\ref{sec:Superluminal-signals-and} we discuss the implications of
superluminal signal propagation for causality. The cosmological scenario of
$k$-essence does not exhibit any causality violations at the classical level,
despite the presence of superluminal signals. Preservation of causality in a
general configuration of $k$-essence field can be viewed as a potential problem,
on the same footing as the chronology protection problem in General Relativity~\cite{Visser:2002ua}.

\section{Physical restrictions on Lagrangians and solutions\label{sec:Physical restrictions}}

In this section we consider some physically necessary restrictions on the possible
Lagrangians $p(X,\phi)$ and solutions $\phi(t)$.

The main physical context for $k$-essence scenarios is the evolution of the
$k$-essence field on the background of a matter component with a fixed equation
of state parameter $w_{m}$. The energy density of $k$-essence is not necessarily
dominant during this evolution. Since $k$-essence scenarios are proposed as
an explanation of the dark energy, we do not consider the case $w_{m}=-1$ (during
primordial inflation, one must also have $w_{m}>-1$ due to the necessity of
the graceful exit). However, we leave open the possibility $w_{m}<-1$.

An important requirement for a field theory is stability. A theory for a field
$\phi$ is stable and ghost-free if the energy density $\varepsilon_{\phi}$
is positive, the speed of sound $c_{s}$ is real (not imaginary), i.e.~$c_{s}^{2}>0$,
and the Lagrangian for linear perturbations has a hyperbolic signature and a
positive sign at the kinetic term. The speed of sound for perturbations on a
given background is given by~\cite{Garriga:1999vw}\begin{equation}
c_{s}^{2}=\frac{p_{,v}}{vp_{,vv}}.\label{eq:speed of sound def}\end{equation}
To obtain the leading terms of the Lagrangian for the perturbations, one writes
a perturbed solution as $\phi=\phi_{0}(t)+\chi(t,\mathbf{x})$ and expands the
Lagrangian $p(X,\phi)$ to second order in $\chi$; the Lagrangian $p(X,\phi)$
is assumed to be an analytic function of $X$ at $X=0$. The relevant terms
are those quadratic in the derivatives of $\chi$,\begin{align}
p(X,\phi) & =p_{,X}\frac{1}{2}\chi_{,\mu}\chi^{,\mu}+\frac{1}{2}p_{,XX}\chi_{,\mu}\chi_{,\nu}\phi_{0}^{,\mu}\phi_{0}^{\nu}+...\nonumber \\
 & \equiv\frac{1}{2}G^{\mu\nu}\chi_{,\mu}\chi_{,\nu}+...\end{align}
It follows that linear perturbations $\chi$ propagate in the effective metric
\begin{equation}
G^{\mu\nu}\equiv p_{,X}g^{\mu\nu}+p_{,XX}\phi_{0}^{,\mu}\phi_{0}^{,\nu}.\label{eq:effective metric for perturbations}\end{equation}
The no-ghost requirement is that the metric $G^{\mu\nu}$ should have the same
signature as $g^{\mu\nu}$. Regardless of whether the 4-gradient $\phi_{0}^{,\mu}$
is spacelike or timelike,%
\footnote{If $\phi_{0}^{,\mu}$ is null, the metric $G^{\mu\nu}$ will have the correct
signature if $p_{,X}>0$ and $p_{,XX}>0$.%
} the resulting conditions are~\cite{Armendariz-Picon:2005nz}\begin{equation}
p_{,X}=\frac{1}{v}p_{,v}>0,\quad p_{,X}+2Xp_{,XX}=p_{,vv}>0.\end{equation}
In the cosmological context, the field $\phi$ is a function of time $t$ only;
in standard $k$-essence scenarios that we are presently considering, $\phi(t)$
grows monotonically with $t$. Hence, $\phi_{0}^{,\mu}$ is timelike and the
velocity $v\equiv\dot{\phi}$ is positive,\begin{equation}
v\equiv\frac{d\phi}{dt}=\sqrt{2X}>0,\quad\frac{\partial}{\partial X}=\frac{1}{v}\frac{\partial}{\partial v}.\label{eq:velocity def}\end{equation}
 We conclude that a physically reasonable cosmological solution should satisfy
(for $v>0$) the conditions\begin{equation}
vp_{,v}-p>0,\quad p_{,v}>0,\quad p_{,vv}>0.\label{eq:conditions for well behaved solution}\end{equation}
It follows that $p(v,\phi)_{v=0}\leq0$ and that $p(v,\phi)$ is a convex, monotonically
growing function of $v$ at fixed $\phi$ (at least for values of $\phi$ and
$v$ relevant in a cosmological scenario). For factorized Lagrangians $p(v,\phi)=K(\phi)Q(v)$,
we find that $Q(v)$ must be a convex, monotonically growing function of $v$
with $Q(0)\leq0$, and also $Q^{\prime}(v)>0$ and $Q^{\prime\prime}(v)>0$
for all values of $v>0$ that are relevant in a given cosmological scenario. 

Finally, we assume that $K(\phi)$ has monotonic behavior at $\phi\rightarrow\infty$.

\section{Superluminal signals and causality\label{sec:Superluminal-signals-and}}

One of the results of this work is a conclusion that every $k$-essence scenario
based on attractor behavior and a Lagrangian of the form~(\ref{eq:Kessence L})
will include an epoch where the perturbations of the $k$-essence field propagate
superluminally. It is therefore pertinent to discuss the possibility of causality
violations in the presence of superluminal signals.

We first consider small perturbations $\phi_{0}+\delta\phi$ of an arbitrary
background solution $\phi_{0}(x)$ in a Lorentz-invariant, nonlinear field theory.
To first order, the evolution of $\delta\phi$ is described by a linear equation
of the form\begin{equation}
G^{\mu\nu}[\phi_{0}]\nabla_{\mu}\nabla_{\nu}\delta\phi+B^{\mu}[\phi_{0}]\nabla_{\mu}\delta\phi+C[\phi_{0}]\delta\phi=0,\label{eq:G mu nu perturbations}\end{equation}
where the coefficients $G^{\mu\nu}$, $B^{\mu}$, and $C$ are determined by
the Lagrangian and depend on the background solution $\phi_{0}$. Unless Eq.~(\ref{eq:G mu nu perturbations})
is hyperbolic (the matrix $G^{\mu\nu}$ having signature $+--\,-$ or equivalent),
the theory will trivially violate causality: an initial-value (Cauchy) problem
will be ill-posed in any reference frame, and the evolution of perturbations
will be physically unpredictable. Therefore, it is necessary to require that
$G^{\mu\nu}$ have a hyperbolic signature. Background solutions $\phi_{0}$
that lead to a parabolic or an elliptic signature of $G^{\mu\nu}$ even in a
small spacetime domain must be avoided as pathological. The cosmological solution
$\phi_{0}(t)$ used in $k$-essence scenarios will be well-behaved if the conditions~(\ref{eq:conditions for well behaved solution})
hold. Below we assume that $G^{\mu\nu}$ has signature $+--\,-$.

Within a sufficiently small spacetime domain, we may regard $G^{\mu\nu}$, $B^{\mu}$,
and $C$ as constants. Then it is straightforward to derive the dispersion relation\begin{equation}
G^{\mu\nu}k_{\mu}k_{\nu}+\textrm{i}B^{\mu}k_{\mu}+C=0\label{eq:dispersion for plane wave}\end{equation}
for plane wave perturbations $\delta\phi(x)\propto\exp\left[\textrm{i}k_{\mu}x^{\mu}\right]$.
In order to send information ({}``signals'' or {}``sounds'') by means of
a perturbation $\delta\phi(x)$, one needs to create a wave front, that is,
a perturbation with an extremely short wavelength and a high frequency. Thus,
wave fronts propagate along wave vectors $k_{\mu}$ determined by the leading
term in Eq.~(\ref{eq:dispersion for plane wave}),\begin{equation}
G^{\mu\nu}k_{\mu}k_{\nu}=0.\end{equation}
Any wave packet consisting of a superposition of plane waves will propagate
behind the wave front. Therefore, a 4-vector $u^{\mu}$ of signal velocity must
lie within the sound cone,\begin{equation}
G^{\mu\nu}u_{\mu}u_{\nu}>0.\label{eq:sound cone def}\end{equation}
Since the {}``sound metric'' $G^{\mu\nu}$ is determined by the local behavior
of the background solution $\phi_{0}(x)$, the sound cone may have an arbitrary
relationship with the lightcone $g^{\mu\nu}u_{\mu}u_{\nu}=0$ determined by
the spacetime metric $g^{\mu\nu}$. Thus, in some theories the sound signal
worldlines may be timelike, null, or even spacelike depending on the spatial
direction of their propagation. 

The speed of sound waves is therefore direction-dependent. The background tensor
$G^{\mu\nu}$ determines (a class of) preferred reference frames where $G^{\mu\nu}$
is diagonal. Propagation of sound is most conveniently described in terms of
sound speeds in different directions in a preferred frame. In this sense, one
may say that a dynamical Lorentz violation takes place for sound waves, although
the full theory (that includes the tensor $G^{\mu\nu}$ as a dynamical quantity)
of course remains Lorentz-invariant.

In the context of $k$-essence cosmology, the sound metric $G^{\mu\nu}$ is
given by Eq.~(\ref{eq:effective metric for perturbations}). Preferred frames
are those where the $t$ axis coincides with the cosmological time. So in the
preferred frames $\phi_{0}=\phi_{0}(t)$ is a function of time only, and the
dispersion relation is\begin{equation}
\omega^{2}=c_{s}^{2}\left|\mathbf{k}\right|^{2},\end{equation}
where $\mathbf{k}$ is the 3-dimensional wave vector and $c_{s}$ is the (direction-independent)
speed of sound defined by Eq.~(\ref{eq:speed of sound def}). In this paper
we show that the considerations of the {}``no-go theorem''~\cite{Bonvin:2006vc}
hold for those Lagrangians of the form~(\ref{eq:Kessence L}) that admit scenarios
of tracking $k$-essence. By virtue of this theorem, there exists an epoch with
$c_{s}^{2}>1$. During this epoch (which may be quite short~\cite{Bonvin:2006vc}),
it is possible to send signals along spacelike worldlines.

If spacelike sound signals propagated in arbitrary \emph{spacetime} directions,
one could easily create closed worldlines made of signals, called {}``closed
signal curves'' (CSCs) in Ref.~\cite{Bonvin:2007mw}. This would open a Pandora's
box of classical time travel paradoxes, also violating the unitarity of quantum
theory (see e.g.~\cite{Fewster:1994gt,Rosenberg:1997ze,Everett:2004ka,Moldoveanu:2007}).
However, the allowed sound signal directions are only those within the sound
cone~(\ref{eq:sound cone def}). This limitation precludes the possibility
of constructing CSCs within a small domain where $G^{\mu\nu}\approx\textrm{const}$.
This can be shown as follows. Diagonalizing the tensor $G^{\mu\nu}$ within
that domain, one finds a preferred reference frame $\left\{ t,x,y,z\right\} $
where sound signals (whether spacelike, null, or timelike) always propagate
in the positive direction along the $t$ axis. Signals sent by conventional
means also propagate in the positive $t$ direction. Since the local coordinates
$\left\{ t,x,y,z\right\} $ are valid within the entire domain where $G^{\mu\nu}\approx\textrm{const}$,
no CSCs are possible within that domain.

It is straightforward to see that no causality violations through CSCs can occur
in $k$-essence cosmology. Since $\dot{\phi}_{0}>0$ at all times, the 4-vector
$\nabla_{\mu}\phi_{0}$ is everywhere timelike and selects a global preferred
reference frame. (Even if $\dot{\phi}_{0}=0$ momentarily, the preferred frame
is still selected by continuity.) In this reference frame, the sound waves propagate
in the direction of increasing coordinate $t$. Hence, there exists a global
foliation of the entire spacetime by spacelike hypersurfaces of equal $t$.
Any sound signals (whether spacelike, null, or timelike), as well as any signals
sent by conventional means, will traverse these hypersurfaces in the direction
of increasing $t$. It follows that CSCs cannot occur, either locally or globally.

Similar conclusions were reached in models of inflation having $c_{s}^{2}>1$~\cite{Mukhanov:2005bu}
as well as in situations involving a $k$-essence field on a black hole background~\cite{Babichev:2006vx,Babichev:2007wg}.
By itself, a superluminal speed of sound does not automatically lead to CSCs
or causality violations.

In certain field theories, one can construct backgrounds $\phi_{0}(x)$ where
CSCs are possible; a notable example is given in Ref.~\cite{Adams:2006sv}.
However, such backgrounds are artificial in the sense that they require an ad
hoc configuration of the field $\phi_{0}(x)$. It remains to be seen whether
such causality-violating backgrounds can occur as a result of the dynamical
evolution of the field $\phi_{0}(x)$ in a cosmological context.

The problem of causality violation by CSCs is similar to the problem of closed
timelike curves (CTCs) occurring in General Relativity~\cite{Visser:2002ua}.
It is difficult to find a metric $g_{\mu\nu}$ that is initially well-behaved
but admits CTCs as a result of dynamical evolution (one such example is given
in Ref.~\cite{Ori:2007kk}). Hawking's {}``chronology protection conjecture''
states that such spacetimes containing CTCs will be always unstable due to quantum
effects; but it remains an open conjecture~\cite{Visser:2002ua}. Similar considerations
apply to CSCs occurring in nonlinear field theories. It is possible that CSCs
will always lead to quantum instabilities due to a similar {}``chronology protection''
mechanism. Further work is needed to resolve this intriguing question.

\section{Equations of motion}

We begin by writing the well-known evolution equations for $k$-essence cosmology
in a convenient set of variables. The equations in this section will be used
at various points in the following analysis.

We consider a spatially flat FRW universe with the metric\begin{equation}
g_{\mu\nu}dx^{\mu}dx^{\nu}=dt^{2}-a^{2}(t)\left[dx^{2}+dy^{2}+dz^{2}\right],\end{equation}
where $a(t)$ is the scale factor. In the epoch of interest, the universe contains
the dynamical $k$-essence field $\phi(t)$ and a matter component with energy
density $\varepsilon_{m}$ and pressure $p_{m}$. The matter component can be
approximately treated as nondynamical in the sense that its equation of state
is fixed,\begin{equation}
w_{m}\equiv\frac{p_{m}}{\varepsilon_{m}}=\textrm{const}.\end{equation}
The energy-momentum tensor of the field $\phi$ is that of a perfect fluid with
pressure $p(X,\phi)$ and energy density \begin{equation}
\varepsilon_{\phi}=2Xp_{,X}-p.\end{equation}
Here and below we denote partial derivatives by a comma, so $p_{,X}\equiv\partial p/\partial X$.
We introduce the velocity $v\equiv\dot{\phi}$ as shown by Eq.~(\ref{eq:velocity def}).
Note that the Lagrangian $p(X,\phi)$ is an analytic function of $X$ and thus
an analytic function of $v^{2}$. 

The equation of state parameter for $k$-essence, $w_{\phi}$, is defined by
\begin{equation}
w_{\phi}\equiv\frac{p(X,\phi)}{\varepsilon_{\phi}}=\frac{p}{vp_{,v}-p}.\end{equation}
 A factorizable Lagrangian~(\ref{eq:Kessence L}) is expressed as a function
of $v$ and $\phi$ as follows,\begin{equation}
p(X,\phi)=K(\phi)Q(v),\quad Q(v)\equiv L(X).\label{eq:Kessence L2}\end{equation}
 For a Lagrangian of this form, $w_{\phi}$ is a function of $v$ only,\begin{equation}
w_{\phi}(v)=\frac{Q}{vQ'-Q},\end{equation}
since the energy density factorizes,\begin{equation}
\varepsilon_{\phi}=K(\phi)\tilde{\varepsilon}_{\phi}(v),\quad\tilde{\varepsilon}_{\phi}(v)\equiv vQ'(v)-Q(v).\end{equation}
We assume that the functions $K$ and $Q$ in Eq.~(\ref{eq:Kessence L2}) are
chosen such that $K(\phi)>0$.

The cosmological evolution is described by the equations of motion for $\phi(t)$,
$\varepsilon_{m}(t)$, and $a(t)$,\begin{align}
\frac{\dot{a}}{a}\equiv H & =\kappa\sqrt{\varepsilon_{\phi}+\varepsilon_{m}},\qquad\kappa^{2}\equiv\frac{8\pi G}{3},\label{eq:adot etc.}\\
\frac{d}{dt}\left(p_{,v}(v,\phi)\right) & \equiv\ddot{\phi}p_{,vv}+\dot{\phi}p_{,\phi v}=-3Hp_{,v}+p_{,\phi},\\
\dot{\varepsilon}_{m} & =-3H\left(\varepsilon_{m}+p_{m}\right)=-3H\left(1+w_{m}\right)\varepsilon_{m}.\label{eq:epsilon m EOM}\end{align}
The equation of motion for the field $\phi$ can be also rewritten as a conservation
law,\begin{equation}
\dot{\varepsilon}_{\phi}=-3H\left(\varepsilon_{\phi}+p(X,\phi)\right)=-3H\left(1+w_{\phi}\right)\varepsilon_{\phi}.\label{eq:epsilon EOM}\end{equation}
The total energy density $\varepsilon_{\text{tot}}\equiv\varepsilon_{\phi}+\varepsilon_{m}$
satisfies the equation\begin{equation}
\dot{\varepsilon}_{\text{tot}}=-3H\varepsilon_{\text{tot}}\left[\left(1+w_{m}\right)+\frac{\varepsilon_{\phi}}{\varepsilon_{\text{tot}}}\left(w_{\phi}-w_{m}\right)\right].\label{eq:epsilon total equ}\end{equation}

Since the equations of motion (\ref{eq:adot etc.})--(\ref{eq:epsilon m EOM})
do not depend explicitly on time, and since $\phi(t)$ is monotonic in $t$,
we may use the value of $\phi$ as the time variable instead of $t$. Then we
obtain a closed system of two first-order equations for $v(\phi)$ and $\varepsilon_{m}(\phi)$,\begin{align}
\frac{dv(\phi)}{d\phi} & =-\frac{vp_{,v\phi}-p_{,\phi}+3\kappa p_{,v}\sqrt{\varepsilon_{m}+vp_{,v}-p}}{vp_{,vv}},\label{eq:v equ}\\
\frac{d\varepsilon_{m}}{d\phi} & =-\frac{3\kappa\left(1+w_{m}\right)\varepsilon_{m}}{v}\sqrt{\varepsilon_{m}+vp_{,v}-p}.\label{eq:epsilon m equ}\end{align}

We will make extensive use of the auxiliary quantity $R$ defined by\begin{equation}
R\equiv\frac{\varepsilon_{m}}{\varepsilon_{\phi}+\varepsilon_{m}}.\end{equation}
 Since energy densities $\varepsilon_{\phi}$ and $\varepsilon_{m}$ are always
positive, the ratio $R$ always remains between 0 and 1. The equation of motion
for $R(\phi)$ is straightforwardly derived from Eqs.~(\ref{eq:epsilon m EOM})--(\ref{eq:epsilon EOM})
and can be written as\begin{equation}
\frac{dR}{d\phi}=-\frac{3H}{v}R(1-R)\left(w_{m}-w_{\phi}(v,\phi)\right).\label{eq:R equ 0}\end{equation}
We may reformulate the equations of motion~(\ref{eq:v equ})--(\ref{eq:epsilon m equ})
as a closed system of equations involving only the variables $v(\phi)$ and
$R(\phi)$. Since \begin{equation}
\varepsilon_{\phi}+\varepsilon_{m}=\frac{\varepsilon_{\phi}}{1-R}=\frac{vp_{,v}-p}{1-R},\end{equation}
 we obtain \begin{align}
\frac{dv}{d\phi} & =-\frac{1}{vp_{,vv}}\left[vp_{,v\phi}-p_{,\phi}+3\kappa p_{,v}\sqrt{\frac{vp_{,v}-p}{1-R}}\right],\\
\frac{dR}{d\phi} & =-\frac{3\kappa}{v}R\sqrt{1-R}\sqrt{vp_{,v}-p}\left(w_{m}-\frac{p}{vp_{,v}-p}\right).\end{align}
For Lagrangians of the form~(\ref{eq:Kessence L2}), these equations are rewritten
as\begin{align}
\frac{dv}{d\phi} & =-c_{s}^{2}(v)\left[\frac{\left(\ln K\right)_{,\phi}v}{1+w_{\phi}(v)}+3\kappa\sqrt{\frac{K(\phi)\tilde{\varepsilon}_{\phi}(v)}{1-R}}\right],\label{eq:v equ clever}\\
\frac{dR}{d\phi} & =-\frac{3\kappa}{v}R\sqrt{1-R}\sqrt{K(\phi)\tilde{\varepsilon}_{\phi}(v)}\left(w_{m}-w_{\phi}(v)\right).\label{eq:R EOM}\end{align}
Here $\tilde{\varepsilon}_{\phi}(v)$, $c_{s}^{2}(v)$, and $w_{\phi}(v)$ are
understood as fixed functions of $v$,\begin{equation}
\tilde{\varepsilon}_{\phi}(v)\equiv vQ'-Q,\quad c_{s}^{2}(v)\equiv\frac{Q'}{vQ''},\quad w_{\phi}(v)\equiv\frac{Q(v)}{\tilde{\varepsilon}_{\phi}(v)},\end{equation}
determined by the given Lagrangian $p(v,\phi)=Q(v)K(\phi)$. These functions
satisfy the following equations,\begin{align}
\frac{d}{dv}\tilde{\varepsilon}_{\phi}(v) & =\frac{1+w_{\phi}(v)}{vc_{s}^{2}(v)}\tilde{\varepsilon}_{\phi}(v),\\
\frac{d}{dv}w_{\phi}(v) & =\frac{1+w_{\phi}(v)}{v}\left[1-\frac{w_{\phi}(v)}{c_{s}^{2}(v)}\right].\label{eq:derivative of wk}\end{align}

\section{Viable Lagrangians for tracking solutions}


The detailed analysis of asymptotically stable solutions is given in Appendix~\ref{sec:Asymptotically-stable-solutions}.
Each asymptotically stable solution is characterized by the asymptotic values
of $v=\dot{\phi}$ and $R=\varepsilon_{m}/\varepsilon_{\text{tot}}$, considered
as functions of $\phi$:\begin{equation}
v_{0}\equiv\lim_{\phi\rightarrow\infty}v(\phi),\quad R_{0}\equiv\lim_{\phi\rightarrow\infty}\frac{\varepsilon_{m}(\phi)}{\varepsilon_{\text{tot}}(\phi)}.\end{equation}
 As a summary of the results, we list all of the possibilities, together with
the requirements on the Lagrangian $p=K(\phi)Q(v)$ and the allowed values of
$v_{0}$, $R_{0}$, $w_{m}$, and $w_{\phi}(v_{0})$. {[}Note that the function
$K(\phi)$ can be always multiplied by a constant, to be absorbed in $Q(v)$.{]}
The requirements listed are necessary and sufficient conditions for the asymptotic
stability of tracker solutions. The applicability of these tracker scenarios
to $k$-essence cosmology is analyzed in subsections~\ref{sub:Radiation-dominated-era},
\ref{sub:Dust-dominated-era}, and \ref{sub:Viable-scenarios}.

\subsection{Tracker solutions\label{sub:Tracker-solutions}\label{sec:Viable-Lagrangians-for}}

\paragraph*{Case 1.}

The function $K(\phi)$ is of the form \begin{equation}
K(\phi)=\frac{1+K_{0}(\phi)}{\phi^{2}},\quad\lim_{\phi\rightarrow\infty}K_{0}(\phi)=0.\end{equation}
The value $v_{0}$ is determined from $w_{\phi}(v_{0})=w_{m}$, and then $R_{0}$
is given by\begin{equation}
R_{0}=1-\frac{9\kappa^{2}Q'(v_{0})^{2}}{4\tilde{\varepsilon}_{\phi}(v_{0})}.\end{equation}
This value of $R_{0}$ must satisfy $0<R_{0}<1$ (the possibility $R_{0}=0$
is equivalent to case 2). The conditions\begin{equation}
v_{0}\neq0,\; c_{s}^{2}(v_{0})\neq0,\;\left|w_{m}\right|<1,\; w_{m}<c_{s}^{2}(v_{0}),\;\tilde{\varepsilon}_{\phi}(v_{0})\neq0\end{equation}
must hold.

\paragraph*{Case 2. }

The function $K(\phi)$ is of the form \begin{equation}
K(\phi)=\frac{1+K_{0}(\phi)}{\phi^{2}},\quad\lim_{\phi\rightarrow\infty}K_{0}(\phi)=0.\end{equation}
The value $v_{0}$ is determined from\begin{equation}
3\kappa\sqrt{\tilde{\varepsilon}_{\phi}(v_{0})}=\frac{2v_{0}}{1+w_{\phi}(v_{0})}\label{eq:case 2 cond}\end{equation}
and must satisfy $v_{0}\neq0$. The following conditions must hold,\begin{equation}
w_{\phi}(v_{0})<w_{m},\;\left|w_{\phi}(v_{0})\right|<1,\; c_{s}^{2}(v_{0})\neq0.\end{equation}
The tracker solution has $R_{0}=0$ ($k$-essence dominates at late times).

\paragraph*{Case 3.}

The function $K(\phi)$ is of the form \begin{equation}
K(\phi)=\frac{K_{0}(\phi)}{\phi^{\alpha}},\quad\lim_{\phi\rightarrow\infty}\frac{\ln K_{0}(\phi)}{\ln\phi}=0,\end{equation}
i.e.~the function $K_{0}$ either tends to a constant, or grows or decays slower
than any power of $\phi$ at $\phi\rightarrow\infty$. This condition determines
the value of $\alpha$. This value of $\alpha$ must satisfy\begin{equation}
2<\alpha<1+\frac{2}{1+w_{m}}.\end{equation}
The interval for $\alpha$ is nonempty if\begin{equation}
\left|w_{m}\right|<1.\end{equation}
The value of $\alpha$ determines $v_{0}$ by\begin{equation}
\alpha=2\frac{1+w_{\phi}(v_{0})}{1+w_{m}}.\end{equation}
The resulting value of $v_{0}$ must satisfy the conditions\begin{equation}
v_{0}\neq0,\; c_{s}^{2}(v_{0})>w_{\phi}(v_{0})>w_{m},c_{s}^{2}(v_{0})\neq0,\;\tilde{\varepsilon}_{\phi}(v_{0})\neq0.\end{equation}
The tracker solution has $R_{0}=1$ ($k$-essence is negligible).

\paragraph*{Case 4.}

The function $K(\phi)$ is of the form \begin{equation}
K(\phi)=\frac{K_{0}(\phi)}{\phi^{2}},\end{equation}
where the function $K_{0}$ must satisfy\begin{equation}
\lim_{\phi\rightarrow\infty}K_{0}(\phi)=0,\quad\lim_{\phi\rightarrow\infty}\frac{\ln K_{0}(\phi)}{\ln\phi}=0,\end{equation}
i.e.~$K_{0}(\phi)$ decays slower than any power of $\phi$ at $\phi\rightarrow\infty$.
The value of $v_{0}$ is determined from the conditions \begin{equation}
w_{\phi}(v_{0})=w_{m},\quad\left|w_{m}\right|<1.\end{equation}
The following conditions must then hold,\begin{equation}
v_{0}\neq0,\; c_{s}^{2}(v_{0})>w_{m},\;\tilde{\varepsilon}_{\phi}(v_{0})\neq0,\; c_{s}^{2}(v_{0})\neq0.\end{equation}
The tracker solution has $R_{0}=1$ ($k$-essence is negligible).

\paragraph*{Case 5.}

The function $K(\phi)$ decays slower than $\phi^{-\alpha}$ (or grows), where\begin{equation}
\alpha\equiv\frac{2}{1+w_{m}},\quad-1<w_{m}<0.\end{equation}
More precisely, \begin{equation}
K(\phi)=\frac{K_{0}(\phi)}{\phi^{\alpha}},\quad\lim_{\phi\rightarrow\infty}K_{0}(\phi)=\infty.\end{equation}
The value of $v_{0}$ is determined as a root of $Q(v_{0})=0$ and $Q^{\prime}(v_{0})=0$,
i.e.~we must have a Taylor expansion near $v=v_{0}$ of the form\begin{equation}
Q(v)=\frac{Q_{0}}{nv_{0}}(v-v_{0})^{n},\quad n\ge2,\; Q_{0}>0.\end{equation}
Then the tracker solution has $R_{0}=1$ ($k$-essence is negligible) and $w_{\phi}(v_{0})=0$.
The value of $v$ must be above $v_{0}$ at all times (or else $c_{s}^{2}<0$).

\paragraph*{Case 6.}

The function $Q(v)$ has an expansion at $v=0$ of the form\begin{equation}
Q(v)=Q_{1}v^{n}+o(v^{n}),\quad Q_{1}>0,\; n>2.\end{equation}
This determines the value of $n$. The function $K(\phi)$ decays slower than
$\phi^{-\alpha}$ (or grows), where\begin{equation}
\alpha\equiv\frac{2n}{\left(n-1\right)\left(1+w_{m}\right)}.\end{equation}
More precisely, \begin{equation}
K(\phi)=\frac{K_{0}(\phi)}{\phi^{\alpha}},\quad\lim_{\phi\rightarrow\infty}K_{0}(\phi)=\infty.\end{equation}
The condition \begin{equation}
-\frac{n-3}{n-1}<w_{m}<\frac{1}{n-1}\label{eq:condition case 6}\end{equation}
must hold. Then the tracker solution has $R_{0}=1$ ($k$-essence is negligible),
$v_{0}=0$, $w_{\phi}(v_{0})=\frac{1}{n-1}$, and $c_{s}^{2}(v_{0})=\frac{1}{n-1}$.

\paragraph*{Case 7.}

The function $K(\phi)$ has the form \begin{equation}
K(\phi)=\frac{K_{0}(\phi)}{\phi^{2}},\end{equation}
where the function $K_{0}(\phi)$ is such that\begin{equation}
\lim_{\phi\rightarrow\infty}K_{0}(\phi)>\frac{1}{9\kappa^{2}Q_{1}}\quad\textrm{or}\quad\lim_{\phi\rightarrow\infty}K_{0}(\phi)=\infty.\end{equation}
The function $Q(v)$ has an expansion at $v=0$ of the form\begin{equation}
Q(v)=Q_{1}v^{2}+o(v^{2}),\quad Q_{1}>0.\end{equation}
We must have $w_{m}>1$. The tracker solution has $R_{0}=0$ ($k$-essence dominates),
$v_{0}=0$, and $w_{\phi}(v_{0})=c_{s}^{2}(v_{0})=1$.

\paragraph*{Case 8.}

The function $K(\phi)$ decays slower than $\phi^{-2}$ or grows,\begin{equation}
K(\phi)=\frac{K_{0}(\phi)}{\phi^{2}},\quad\lim_{\phi\rightarrow\infty}K_{0}(\phi)=\infty.\end{equation}
The value of $v_{0}$ is determined from $Q^{\prime}(v_{0})=0$, $Q(v_{0})<0$.
More precisely, we have an expansion near $v=v_{0}$,\begin{equation}
Q(v)=Q_{0}+Q_{2}(v-v_{0})^{n},\quad Q_{0}<0,\quad n\ge2.\end{equation}
We must have $v_{0}\neq0$ and $w_{m}>-1$. The tracker solution has $R_{0}=0$
($k$-essence dominates) and $w_{\phi}(v_{0})=-1$. The value of $v$ must be
above $v_{0}$ at all times (or else $c_{s}^{2}<0$).

\paragraph*{Case 9.}

The function $K(\phi)$ decays slower than $\phi^{-2}$ or grows,\begin{equation}
K(\phi)=\frac{K_{0}(\phi)}{\phi^{2}},\quad\lim_{\phi\rightarrow\infty}K_{0}(\phi)=\infty.\end{equation}
The value of $v_{0}$ is determined from $Q^{\prime}(v_{0})=0$, $Q(v_{0})=0$.
More precisely, we have an expansion near $v=v_{0}$,\begin{equation}
Q(v)=Q_{1}(v-v_{0})^{n},\quad n\ge2.\end{equation}
We must have $w_{m}>0$ and $v_{0}\neq0$. The tracker solution has $R_{0}=0$
($k$-essence dominates) and $w_{\phi}(v_{0})=0$. The value of $v$ must be
above $v_{0}$ at all times (or else $c_{s}^{2}<0$).

\paragraph*{Case 10. }

The function $K(\phi)$ decays slower than $\phi^{-2}$ or grows,\begin{equation}
K(\phi)=\frac{K_{0}(\phi)}{\phi^{2}},\quad\lim_{\phi\rightarrow\infty}K_{0}(\phi)=\infty.\end{equation}
The function $Q(v)$ must have an expansion near $v=0$ of the form\begin{equation}
Q(v)=-Q_{0}+Q_{1}v^{n},\quad Q_{0}>0,\; n\ge2.\end{equation}
We must have $w_{m}>-1$. The tracker solution has $R_{0}=0$ ($k$-essence
dominates), $v_{0}=0$, and $w_{\phi}(v_{0})=-1$.

\paragraph*{Case 11.}

The function $K(\phi)$ decays slower than $\phi^{-2}$ or grows,\begin{equation}
K(\phi)=\frac{K_{0}(\phi)}{\phi^{2}},\quad\lim_{\phi\rightarrow\infty}K_{0}(\phi)=\infty.\end{equation}
The function $Q(v)$ must have an expansion near $v=0$ of the form\begin{equation}
Q(v)=Q_{1}v^{n}+o(v^{n}),\quad Q_{1}>0,\; n>2.\end{equation}
This determines the value of $n$. The condition\begin{equation}
w_{m}>\frac{1}{n-1}\end{equation}
 must hold. The tracker solution has $R_{0}=0$ ($k$-essence dominates), $v_{0}=0$,
and $w_{\phi}(v_{0})=c_{s}^{2}=\frac{1}{n-1}$.

\paragraph*{Case 12.}

The function $Q(v)$ must have an expansion near $v=0$ of the form\begin{equation}
Q(v)=Q_{1}v^{n}+Q_{2}v^{n+p},\quad Q_{1}>0,\; n>2,\; p>0.\end{equation}
This determines the values of $n$ and $p$. The function $K(\phi)$ must be
of the form\begin{equation}
K(\phi)=\frac{K_{0}(\phi)}{\phi^{2}},\end{equation}
where $K_{0}(\phi)$ must satisfy\begin{equation}
\lim_{\phi\rightarrow\infty}K_{0}(\phi)=\infty,\quad\int^{\infty}\frac{d\phi}{\phi}K_{0}^{-\frac{p}{n-2}}(\phi)=\infty.\end{equation}
(The function $K_{0}(\phi)$ grows slower than $\left(\ln\phi\right)^{(n-2)/p}$.)
We must have $w_{m}=\frac{1}{n-1}$. The tracker solution has $R_{0}=0$ ($k$-essence
dominates), $v_{0}=0$, and $w_{\phi}(v_{0})=c_{s}^{2}=\frac{1}{n-1}$.

\subsection{Radiation-dominated era\label{sub:Radiation-dominated-era}}

We now select Lagrangians that admit tracker solutions during radiation domination,
$w_{m}=\frac{1}{3}$. In order not to violate the nucleosynthesis bound, the
energy density of $k$-essence must be subdominant throughout the radiation
era~\cite{Armendariz-Picon:2000ah}, \begin{equation}
R_{0}\gtrsim0.99.\label{eq: R_0_bound}\end{equation}
 Admissible trackers may have a value $R_{0}$ within the range $0.99\lesssim R_{0}<1$,
or $R_{0}=1$. A solution with $0<R_{0}<1$ is only possible with Lagrangians
given by case 1,\begin{equation}
K(\phi)=\frac{1+K_{0}(\phi)}{\phi^{2}},\quad\lim_{\phi\rightarrow\infty}K_{0}(\phi)=0.\end{equation}
We denote by $v_{r}$ the asymptotic value of $v$ during the radiation era.
Possible values of $v_{r}$ are determined from $w_{\phi}(v_{r})=\frac{1}{3}$,
and $v_{r}$ must satisfy \begin{equation}
c_{s}^{2}(v_{r})>\frac{1}{3},\quad\tilde{\varepsilon}_{\phi}(v_{r})\neq0,\quad v_{r}\neq0.\end{equation}
The corresponding value of $R_{0}$ must respect the bound~(\ref{eq: R_0_bound}),\begin{equation}
R_{0}=1-\left.\frac{9\kappa^{2}Q^{\prime2}}{4\tilde{\varepsilon}_{\phi}}\right|_{v=v_{r}}\gtrsim0.99.\end{equation}

Solutions with $R_{0}=1$ and $w_{m}=\frac{1}{3}$ are possible in cases 3,
4, and 6. The first set of solutions is given by\begin{equation}
K(\phi)=\frac{K_{0}(\phi)}{\phi^{\alpha}},\quad\lim_{\phi\rightarrow\infty}\frac{\ln K_{0}(\phi)}{\ln\phi}=0,\end{equation}
where $2<\alpha<\frac{5}{2}$. Admissible functions $K_{0}(\phi)$ decay or
grow slower than any power of $\phi$, e.g.~$K_{0}(\phi)\propto(\ln\phi)^{\beta}$.
Admissible values of $v_{r}$ are determined from the conditions

\begin{equation}
w_{\phi}(v_{r})=\frac{2\alpha}{3}-1,\quad\tilde{\varepsilon}_{\phi}(v_{r})\neq0,\quad v_{r}\neq0.\end{equation}

The second set of Lagrangians is\begin{equation}
K(\phi)=\frac{K_{0}(\phi)}{\phi^{2}},\quad\lim_{\phi\rightarrow\infty}K_{0}(\phi)=0,\quad\lim_{\phi\rightarrow\infty}\frac{\ln K_{0}(\phi)}{\ln\phi}=0.\end{equation}
 The possible values of $v_{r}$ are determined from $w_{\phi}(v_{r})=\frac{1}{3}$,
and the following conditions must be also satisfied, \begin{equation}
c_{s}^{2}(v_{r})>\frac{1}{3},\quad\tilde{\varepsilon}_{\phi}(v_{r})\neq0,\quad v_{r}\neq0.\end{equation}

The third set of admissible Lagrangians is described by case 6 with $n=3$,
namely\begin{align}
K(\phi) & =\frac{K_{0}(\phi)}{\phi^{9/4}},\quad\lim_{\phi\rightarrow\infty}K_{0}(\phi)=\infty,\\
Q(v) & =Q_{1}v^{3}+o(v^{3}),\quad Q_{1}>0.\end{align}
In this case, $v_{r}=0$. The solution of case 6 with $n\geq4$ cannot be used
since the condition~(\ref{eq:condition case 6}) cannot be satisfied with $w_{m}=\frac{1}{3}$.

\subsection{Dust-dominated era\label{sub:Dust-dominated-era}}

We now select the tracker solutions that exist for $w_{m}=0$. In order to describe
the late-time domination of $k$-essence, we must look for solutions with $w_{\phi}<-\frac{1}{3}$
and $R_{0}=0$. The possible trackers are cases 2, 8, and 10.

In case 2, the Lagrangian must satisfy\begin{equation}
K(\phi)=\frac{1+K_{0}(\phi)}{\phi^{2}},\quad\lim_{\phi\rightarrow\infty}K_{0}(\phi)=0.\end{equation}
We denote by $v_{d}$ the asymptotic value of $v$ during the dust era. The
admissible values of $v_{d}\neq0$ are determined from \begin{equation}
3\kappa\sqrt{\tilde{\varepsilon}_{\phi}(v_{d})}=\frac{2v_{d}}{1+w_{\phi}(v_{d})}.\label{eq:case2 cond A}\end{equation}
In addition, the following conditions must be satisfied:\begin{equation}
-1<w_{\phi}(v_{d})<0,\quad c_{s}^{2}(v_{d})\neq0.\label{eq:case2 cond B}\end{equation}

The second set of Lagrangians is for cases 8 and 10,\begin{equation}
K(\phi)=\frac{K_{0}(\phi)}{\phi^{2}},\quad\lim_{\phi\rightarrow\infty}K_{0}(\phi)=\infty.\end{equation}
This condition for $K(\phi)$ is satisfied, for example, by $K(\phi)\propto\phi^{\alpha}$
with $\alpha>-2$. The value $v_{d}$ must be such that \begin{equation}
Q(v_{d})<0,\quad Q'(v_{d})=0,\end{equation}
while we may have either $v_{d}\neq0$ or $v_{d}=0$.

\subsection{Viable scenarios\label{sub:Viable-scenarios}}

Having listed all the Lagrangians that admit desired solutions in the radiation-
and dust-dominated eras, it remains to determine the overlap between these classes
of Lagrangians. By comparing the requirements on the functions $K(\phi)$ and
$Q(v)$, we find only two possibilities for trackers in the radiation/dust era:
case 1/case 2 and case 6/case 8.

The first set of Lagrangians (case 1/case 2) is \begin{equation}
K(\phi)=\frac{1+K_{0}(\phi)}{\phi^{2}},\quad\lim_{\phi\rightarrow\infty}K_{0}(\phi)=0.\label{eq: Familiar}\end{equation}
In the radiation era, the asymptotic value of $v$ is given by $v_{r}\neq0$
such that \begin{equation}
w_{\phi}(v_{r})=\frac{1}{3},\quad c_{s}^{2}(v_{r})>\frac{1}{3},\quad\tilde{\varepsilon}_{\phi}(v_{r})\neq0,\end{equation}
and the dust attractor is given by $v_{d}\neq0$ such that Eqs.~(\ref{eq:case2 cond A})--(\ref{eq:case2 cond B})
hold. These Lagrangians describe the well-known scenario~\cite{Armendariz-Picon:2000dh}
where the $k$-essence tracks radiation during the radiation era and eventually
starts to dominate in the dust era. The function $Q(v)$ must be chosen to satisfy
the conditions of cases 1 and 2. Additionally, one must exclude the possibility
of a dust tracker (case 1, $w_{m}=0$) by adjusting $Q(v)$ such that the conditions
of case 1 are not satisfied for $w_{\phi}(v_{0})=w_{m}=0$~\cite{Armendariz-Picon:2000ah}.

The second set of Lagrangians is described by case 6/case 8. The function $K(\phi)$
is of the form\[
K(\phi)=\frac{K_{0}(\phi)}{\phi^{2}},\quad\lim_{\phi\rightarrow\infty}K_{0}(\phi)=\infty.\]
The function $Q(v)$ must be such that\begin{equation}
Q(v)=Q_{1}v^{3}+o(v^{3}),\quad Q_{1}>0.\end{equation}
Then the asymptotic values of $v$ are $v_{r}=0$ in the radiation era (where
$w_{\phi}\approx\frac{1}{2}$) and $v_{d}\neq0$ in the dust era (where $w_{\phi}\approx-1$).
The value $v_{d}$ must be a root of $Q'(v)$ such that\begin{equation}
Q(v_{d})<0,\quad Q'(v_{d})=0.\end{equation}
This scenario, however, has a fatal flaw. The attractor of case 8 requires that
$v>v_{d}$ at all times, while the attractor of case 6 is realized at very small
$v\approx0$. Therefore, a transition from the first attractor to the second
will necessarily involve values of $v<v_{d}$ for which the theory is unstable
since $c_{s}^{2}(v)<0$. Hence, this scenario must be discarded.

Thus we conclude that successful models of $k$-essence are produced only by
Lagrangians described by Eq.~(\ref{eq: Familiar}) under the conditions of
case 1 and case 2.

\subsection{The existence of a superluminal epoch}

We have shown that the only viable $k$-essence scenario is described by case
1/case 2 of Sec.~\ref{sec:Viable-Lagrangians-for}. Now we demonstrate that
in these scenarios $c_{s}^{2}(v_{*})>1$ for some value $v_{*}$ that is reached
during the dust-dominated epoch. The argument is similar to that in Ref.~\cite{Bonvin:2006vc}.

Since $Q(v_{r})>0$ and $Q(v_{d})<0$, while $Q(v)$ is a monotonically growing
function of $v$, we must have $v_{d}<v_{r}$. In both scenarios of case 1 and
case 2, the asymptotic fraction of the energy density $R_{0}$ is equal to a
certain function $F$ of $v_{0}$,\begin{equation}
R_{0}=F(v_{0})\equiv\left.1-\frac{9\kappa^{2}}{4}\frac{Q^{\prime2}}{vQ'-Q}\right|_{v=v_{0}}.\end{equation}
In case 2, $F(v_{0})=0$ due to Eq.~(\ref{eq:case 2 cond}); therefore, we
may describe both cases 1 and 2 by a single function $F(v_{0})$. We note that
$\tilde{\varepsilon}_{\phi}(v)=vQ'(v)-Q(v)$ is a monotonically growing function
of $v$ because\begin{equation}
\frac{d}{dv}\tilde{\varepsilon}_{\phi}(v)=vQ^{\prime\prime}(v)>0\quad\textrm{for }v>0.\end{equation}
Since $\tilde{\varepsilon}_{\phi}(v)>0$ for every relevant value of $v$, it
follows that $F(v)$ is a continuous function for these $v$. For a successful
model of $k$-essence, the radiation tracker must have $F(v_{r})\gtrsim0.99$
and the dust tracker must have $F(v_{d})=0$. During the evolution from the
first tracker to the second, the value of $v$ must traverse the interval $[v_{d},v_{r}]$.
The condition $F(v_{d})<F(v_{r})$ implies (due to the continuity of $F$) that
there exists a value $v_{1}\in[v_{d},v_{r}]$ such that $F^{\prime}(v_{1})$
is positive:\begin{equation}
F^{\prime}(v_{1})=-\left.\frac{9\kappa^{2}Q'Q''}{2(vQ'-Q)^{2}}\left[\frac{vQ'}{2}-Q\right]\right|_{v=v_{1}}>0.\end{equation}
Since $Q'>0$, $Q''>0$, and $\tilde{\epsilon}_{\phi}=vQ'-Q>0$ for all $v\in[v_{d},v_{r}]$,
we can simplify this condition to\begin{equation}
\left.\frac{vQ'}{2}-Q\right|_{v=v_{1}}<0,\end{equation}
or equivalently to \begin{equation}
w_{\phi}(v_{1})=\left.\frac{Q}{vQ'-Q}\right|_{v=v_{1}}>1.\end{equation}

The equation of state parameter $w_{\phi}(v)$ is a continuous function of $v$
that satisfies \begin{equation}
0=w_{\phi}(v_{d})<1<w_{\phi}(v_{1}).\end{equation}
Hence, there exists a value $v_{*}\in[v_{d},v_{1}]$ such that $w_{\phi}(v_{*})>1$
and $w_{\phi}^{\prime}(v_{*})>0$. 

Finally, we show that $c_{s}^{2}(v_{*})>1$ follows from the conditions $w_{\phi}(v_{*})>1$
and $w_{\phi}^{\prime}(v_{*})>0$. According to Eq.~(\ref{eq:derivative of wk}),
we have\begin{equation}
w_{\phi}^{\prime}(v)=\left.\frac{\left(1+w_{\phi}\right)\left(c_{s}^{2}-w_{\phi}\right)}{vc_{s}^{2}}\right|_{v=v_{*}}>0.\end{equation}
Therefore\begin{equation}
c_{s}^{2}(v_{*})>w_{\phi}(v_{*})>1.\end{equation}
 Since $c_{s}^{2}(v)$ is a continuous function, this demonstrates the existence
of an interval of values of $v$ within $\left[v_{d},v_{r}\right]$ where $c_{s}^{2}(v)>1$.
This superluminal epoch occurs during the dust-dominated era.

\section*{Acknowledgments}

The authors thank Slava Mukhanov and Alex Vikman for useful discussions. Jin
U Kang is supported by the German Academic Exchange Service (DAAD). Vitaly Vanchurin
is supported in part by the project {}``Transregio (Dark Universe).''

\appendix

\section{Asymptotically stable solutions\label{sec:Asymptotically-stable-solutions}}

The standard analysis of the $k$-essence trackers (e.g.~\cite{Armendariz-Picon:2000ah})
involves several simplifying but restrictive assumptions concerning the behavior
of the solutions. A wider range of $k$-essence models will be obtained if some
of these assumptions are lifted. Let us therefore characterize the desired features
of the cosmological evolution of $k$-essence in a general manner.

Scenarios of $k$-essence are based on the assumption that the field $\phi$
has an almost constant equation of state parameter ($w_{\phi}$) during a cosmologically
long epoch while another matter component dominates the energy density of the
universe. Eventually, the $k$-essence itself becomes dominant and plays the
role of {}``dark energy,'' again with an approximately constant $w_{\phi}$.
It is important that the solution curves serve as attractors for all neighbor
solutions. In that case, the value of $w_{\phi}$ at late times is essentially
independent of the initial conditions.

When the radiation-dominated epoch gives way to the epoch of dust domination,
the behavior of $k$-essence will change in a model-dependent way. However,
it is technically convenient to study the behavior of $k$-essence under the
assumption that the dominant matter component has a fixed equation of state
for all time. Then the existence of tracker solutions will be found by studying
the asymptotic behavior of the solutions at $t\rightarrow\infty$ (equivalently,
at $\phi\rightarrow\infty$). This is the approach taken in this paper.

The evolution of $k$-essence together with a single matter component is described
by the equations of motion (EOM) shown above as Eqs.~(\ref{eq:v equ})--(\ref{eq:epsilon m equ})
in terms of the variables $\left\{ v(\phi),\varepsilon_{m}(\phi)\right\} $.
We call a solution $\left\{ v(\phi),\varepsilon_{m}(\phi)\right\} $ \textsl{asymptotically
stable} if $w_{\phi}(\phi)$ tends to a constant at $\phi\rightarrow\infty$
and if \emph{all} neighbor solutions (at least within a finite domain of attraction)
also approach the same value of $w_{\phi}$. In this section, we restrict our
attention to asymptotically stable solutions with one matter component. Since
reasonable values of $w_{\phi}$ are within the interval $\left[-1,1\right]$,
it is justifiable to ignore solutions where $w_{\phi}$ tends to infinity at
late times. In principle, one could also have solutions where $w_{\phi}(\phi)$
oscillates without reaching any limit as $\phi\rightarrow\infty$, but such
solutions are of little physical interest since the value of $w_{\phi}$ at
the end of a given cosmological epoch will then be largely unpredictable. Since
in the models under consideration $w_{\phi}$ is a function of $v$ only, solutions
$v(\phi)$ that oscillate without reaching any limit are also excluded. Applicability
of the effective field theory requires that the derivatives of $\phi$ remain
bounded; thus $v=\dot{\phi}$ cannot diverge to infinity as $\phi\rightarrow\infty$
and must also tend to a constant value, $v(\phi)\rightarrow v_{0}<\infty$. 

There may also exist solutions with initially negligible but growing ratio $\varepsilon_{\phi}/\varepsilon_{m}$.
Such solutions may have a stable behavior with an almost constant $w_{\phi}$
for a finite (but very long) time, until the energy density of $k$-essence
starts to dominate. We do not consider such {}``transient attractors'' in
the present paper.

Our main task is to deduce the possible $k$-essence Lagrangians $p(X,\phi)$
that admit physically meaningful asymptotically stable solutions. We consider
only Lagrangians that have a factorized form~(\ref{eq:Kessence L2}).%
\footnote{Since the analysis uses only the properties of the Lagrangian in the asymptotic
limit $\phi\rightarrow\infty$, our results will apply to more general Lagrangians
that have the form $p=K(\phi)L(X)$ asymptotically at large $\phi$ and fixed
$X$.%
} We assume that the matter component has a constant equation of state parameter
$w_{m}$ such that $w_{m}\neq-1$. 

It will be convenient to use also the auxiliary variable $R(\phi)$ satisfying
the EOM~(\ref{eq:R EOM}). Since the values of $R$ are limited to the interval
$\left[0,1\right]$, any asymptotically stable solution will necessarily approach
a constant value, $R(\phi)\rightarrow R_{0}$ as $\phi\rightarrow\infty$. The
possible values of $R_{0}$ and $v_{0}$ are yet to be determined; the cases
when $R_{0}$ or $v_{0}$ assume critical values ($R_{0}=0$, $R_{0}=1$, $v_{0}=0$)
will need to be treated separately.

The general method of analysis is the following. We have a system of nonlinear
EOM parameterized by a pair of functions $K(\phi),Q(v)$; the general solution
of the EOM is not available in closed form. Our purpose is to determine the
functions $K(\phi),Q(v)$ for which a solution of the EOM exists with the asymptotic
stability property. We first \emph{assume} the existence of an asymptotically
stable solution $\left\{ v(\phi),\varepsilon_{m}(\phi)\right\} $ and derive
the \emph{necessary} conditions on the functions $K(\phi)$ and $Q(v)$ that
admit such solutions (perhaps in more convenient variables, such as $\left\{ v(\phi),R(\phi)\right\} $).
At this step, there will be many cases corresponding to different asymptotic
behavior of $v(\phi)$ and $R(\phi)$. For instance, $v(\phi)$ may tend either
to a nonzero constant or to zero, etc. In each case, we then obtain the \emph{general}
solution of the EOM (with two integration constants) near the assumed stable
solution (e.g.~$v(\phi)=v_{0}-A(\phi)$, with $A(\phi)$ very small). At this
point, it is possible to make simplifying assumptions because we only consider
the solutions in the asymptotic limit $\phi\rightarrow\infty$ and infinitesimally
close to an assumed trajectory. We then investigate whether the general solution
is attracted to the assumed stable solution. In this way, we either obtain \emph{sufficient}
conditions for the existence of a stable solution of an assumed type, or conclude
that no stable solution exists in a given case. After enumerating all the cases,
we will thus obtain necessary and sufficient conditions on $K(\phi)$ and $Q(v)$
for every possible type of stable tracking behavior.

Let us begin by drawing some general consequences about the asymptotic behavior
of stable solutions at $\phi\rightarrow\infty$. Rewriting Eq.~(\ref{eq:epsilon m equ})
as\begin{equation}
\frac{d\varepsilon_{m}^{-1/2}(\phi)}{d\phi}=\frac{3\kappa\left(1+w_{m}\right)}{2v(\phi)\sqrt{R(\phi)}},\label{eq:epsilon m clever 1}\end{equation}
and noting that the right-hand side of Eq.~(\ref{eq:epsilon m clever 1}) is
bounded away from zero, we conclude that $\varepsilon_{m}(\phi)$ decays either
as $\phi^{-2}$ or faster at $\phi\rightarrow\infty$, depending on whether
$v(\phi)\sqrt{R(\phi)}$ tends to zero at large $\phi$. In the following subsections,
we consider all the possible cases. 

Based on the motivation for introducing $k$-essence, we have assumed that $w_{m}\neq-1$.
According to Eq.~(\ref{eq:epsilon m clever 1}), for $w_{m}<-1$ (phantom matter)
the energy density $\varepsilon_{m}$ will satisfy the differential inequality\begin{equation}
\frac{d}{dt}\varepsilon_{m}^{-1/2}=-\frac{3\kappa\left|1+w_{m}\right|}{2\sqrt{R(\phi)}}<-C_{1},\end{equation}
where $C_{1}$ is a positive constant. Thus, $\varepsilon_{m}(t)$ will reach
infinity in finite time regardless of the behavior of $R(\phi)$ and $v(\phi)$.
However, this time can be quite long and the phantom behavior might be only
a temporary phenomenon. Therefore, we will use the property $w_{m}\neq-1$ but
avoid assuming that $w_{m}>-1$.

In the analysis below, we will also use the following elementary facts: 

a) If a function $F(\phi)$ monotonically goes to a constant at $\phi\rightarrow\infty$,
then $F'(\phi)$ decays faster than $\phi^{-1}$. This is easily established
using the identity\begin{equation}
F(0)-\lim_{\phi\rightarrow\infty}F(\phi)=-\int_{0}^{\infty}F'(\phi)d\phi<\infty,\end{equation}
which means that $F'(\phi)$ is integrable at $\phi\rightarrow\infty$. Hence,
$F'(\phi)$ decays faster than $\phi^{-1}$ at $\phi\rightarrow\infty$.

b) If a function $F(\phi)$ is monotonic, then $F'(\phi)\rightarrow0$ if and
only if\begin{equation}
\lim_{\phi\rightarrow\infty}\frac{F(\phi)}{\phi}=0.\end{equation}
This statement follows from the L'Hopital's rule in case $F(\phi)\rightarrow\infty$,
and is trivial in case $F(\phi)$ has a finite limit at $\phi\rightarrow\infty$.

\subsection{Energy density $\varepsilon_{m}\propto\phi^{-2}$ and $R_{0}\neq1$, main case\label{sub:Energy-density-1}}

According to Eq.~(\ref{eq:epsilon m clever 1}), the asymptotic behavior $\varepsilon_{m}\propto\phi^{-2}$
is possible only if $v(\phi)\sqrt{R(\phi)}$ stays bounded away from zero as
$\phi\rightarrow\infty$, in other words if $v_{0}\neq0$ and $R_{0}\neq0$.
We also assume $R_{0}\neq1$, meaning that the energy density of $k$-essence
tracks the matter component; thus $\varepsilon_{\phi}(\phi)\propto\phi^{-2}$
as well. It follows that $H(\phi)\propto\sqrt{\varepsilon_{\phi}(\phi)}\propto\phi^{-1}$,
and then Eq.~(\ref{eq:R equ 0}) yields\begin{equation}
\frac{dR(\phi)}{d\phi}\propto\frac{w_{m}-w_{\phi}(\phi)}{\phi}.\end{equation}
Since $R(\phi)\rightarrow\textrm{const}$, the derivative $dR/d\phi$ must decay
faster than $\phi^{-1}$ as $\phi\rightarrow\infty$. Hence $w_{\phi}(\phi)\rightarrow w_{m}$
as $\phi\rightarrow\infty$. This is the standard tracker behavior: the equation
of state parameters of $k$-essence and matter become almost equal at late times.

By assumption, at large $\phi$ the Lagrangian is factorized, $p=K(\phi)Q(v)$,
and then we have\begin{equation}
w_{m}=w_{\phi}(v_{0})=\frac{Q(v_{0})}{v_{0}Q^{\prime}(v_{0})-Q(v_{0})}.\label{eq:v0 determined}\end{equation}
This algebraic equation determines the possible values of $v_{0}$ for a given
$w_{m}$. (Tracker solutions of this type are impossible if this equation has
no roots.) The property $\varepsilon_{\phi}(\phi)\propto\phi^{-2}$ becomes\begin{equation}
\varepsilon_{\phi}(\phi)=K(\phi)\left(vQ'-Q\right)\propto\phi^{-2}.\end{equation}
Generically one expects \begin{equation}
\tilde{\varepsilon}_{\phi}(v_{0})\equiv v_{0}Q^{\prime}(v_{0})-Q(v_{0})\neq0,\label{eq:epsilon tilde nonzero}\end{equation}
and we temporarily make this additional assumption. Then we obtain \begin{equation}
K(\phi)\propto\phi^{-2}\:\textrm{as}\:\phi\rightarrow\infty.\end{equation}
 This is somewhat more general than the function $K(\phi)=\textrm{const}\cdot\phi^{-2}$
usually considered in $k$-essence models. 

We may consider Lagrangians $p=K(\phi)Q(v)$ with the function $K(\phi)$ of
the form\begin{equation}
K(\phi)=\frac{1+K_{0}(\phi)}{\phi^{2}},\quad\lim_{\phi\rightarrow\infty}K_{0}(\phi)=0.\label{eq:K ansatz 2}\end{equation}
 Let us now derive a sharp condition for the existence of an asymptotically
stable solution $\left\{ v(\phi),R(\phi)\right\} $ in this case. We use the
ansatz \begin{equation}
v(\phi)=v_{0}-A(\phi),\quad R(\phi)=R_{0}-B(\phi),\label{eq:v R ansatz}\end{equation}
where by assumption the unknown functions $A(\phi),B(\phi)$ tend to zero at
$\phi\rightarrow\infty$. After deriving and solving the equations for $A(\phi)$
and $B(\phi)$, we will need to verify this assumption.

Since the left-hand side of Eq.~(\ref{eq:v equ clever}) is $-A'$, it tends
to zero faster than $\phi^{-1}$. On the other hand, assuming that $c_{s}^{2}(v_{0})\neq0$,
we find that the right-hand side of Eq.~(\ref{eq:v equ clever}) contains leading
terms of order $\phi^{-1}$, such as $\left(\ln K\right)_{,\phi}$ and $\sqrt{K}$.
Hence, these terms must cancel, which entails\begin{align}
3\kappa\sqrt{\frac{\tilde{\varepsilon}_{\phi}(v_{0})}{1-R_{0}}} & =\frac{2v_{0}}{1+w_{\phi}(v_{0})}=\frac{2v_{0}}{1+w_{m}}=\frac{2\tilde{\varepsilon}_{\phi}(v_{0})}{Q'(v_{0})}.\label{eq:R0 determined}\end{align}
Since $v_{0}$ is determined from Eq.~(\ref{eq:v0 determined}), this condition
fixes the value of $R_{0}$,\begin{equation}
R_{0}=1-\frac{9\kappa^{2}}{4}\frac{Q^{\prime2}(v_{0})}{v_{0}Q'(v_{0})-Q(v_{0})}.\end{equation}
 The requirement that the values of $R_{0}$ be between $0$ and $1$ further
restricts the possible functions $Q(v)$. Using Eq.~(\ref{eq:R0 determined}),
the condition $R_{0}>0$ can be expressed equivalently as\begin{equation}
Q(v_{0})<\frac{4}{9\kappa^{2}}v_{0}^{2}\frac{w_{m}}{\left(1+w_{m}\right)^{2}}.\end{equation}
No tracker solution is possible if this condition is violated.

The equations for $A(\phi)$ and $B(\phi)$ are now found by linearizing the
equations~(\ref{eq:v equ clever})--(\ref{eq:R EOM}). For brevity, we rewrite
these equations as\begin{align}
\frac{dv}{d\ln\phi} & =-\Lambda_{1}(v)\frac{d\ln K}{d\ln\phi}-\Lambda_{2}(v)\sqrt{\frac{1+K_{0}(\phi)}{1-R}},\label{eq:v equ 2}\\
\frac{dR}{d\ln\phi} & =-\Lambda_{3}(v,R)\sqrt{1+K_{0}(\phi)}\left(w_{m}-w_{\phi}(v)\right),\label{eq:R equ 2}\end{align}
where the auxiliary functions $\Lambda_{1},\Lambda_{2},\Lambda_{3}$ are defined
by\begin{align}
\Lambda_{1}(v) & \equiv\frac{c_{s}^{2}(v)v}{1+w_{\phi}(v)}=\frac{\tilde{\varepsilon}_{\phi}}{vQ^{\prime\prime}(v)},\label{eq:Lambda1 def}\\
\Lambda_{2}(v) & \equiv3\kappa c_{s}^{2}(v)\sqrt{\tilde{\varepsilon}_{\phi}(v)},\label{eq:Lambda2 def}\\
\Lambda_{3}(v,R) & \equiv\frac{3\kappa}{v}R\sqrt{1-R}\sqrt{\tilde{\varepsilon}_{\phi}(v)}=\frac{R\sqrt{1-R}}{vc_{s}^{2}(v)}\Lambda_{2}(v).\end{align}
 Note that Eq.~(\ref{eq:R0 determined}) is equivalent to \begin{equation}
2\Lambda_{1}(v_{0})=\frac{\Lambda_{2}(v_{0})}{\sqrt{1-R_{0}}}.\label{eq:Lambda 1 2 identity}\end{equation}
Substituting the ansatz~(\ref{eq:v R ansatz}) into Eqs.~(\ref{eq:v equ 2})--(\ref{eq:R equ 2}),
using the identity~(\ref{eq:Lambda 1 2 identity}), and keeping only the leading
linear terms, we find\begin{align}
\frac{dA}{d\ln\phi} & =\left(\phi K_{0}^{\prime}+K_{0}\right)\Lambda_{1}(v_{0})-\alpha_{0}A-\frac{\Lambda_{1}(v_{0})B}{1-R_{0}},\label{eq:dA dlnphi}\\
\frac{dB}{d\ln\phi} & =\Lambda_{3}(v_{0},R_{0})w_{\phi}^{\prime}(v_{0})A,\label{eq:dB dlnphi}\end{align}
where we defined the auxiliary constant $\alpha_{0}$ by\begin{equation}
\alpha_{0}\equiv\frac{\Lambda_{2}^{\prime}(v_{0})}{\sqrt{1-R_{0}}}-2\Lambda_{1}^{\prime}(v_{0})=\frac{1-w_{m}}{1+w_{m}}.\end{equation}
For the moment, we assume additionally that\begin{equation}
w_{\phi}^{\prime}(v_{0})\equiv\left(\frac{Q}{vQ'-Q}\right)_{v=v_{0}}^{\prime}=\left(1-\frac{w_{m}}{c_{s}^{2}(v_{0})}\right)\frac{1+w_{m}}{v_{0}}\neq0.\label{eq:w prime nonzero}\end{equation}
Differentiating Eq.~(\ref{eq:dB dlnphi}) with respect to $\ln\phi$ and substituting
into Eq.~(\ref{eq:dA dlnphi}), we find a closed second-order equation for
$A(\phi)$,\begin{equation}
\frac{d^{2}B}{d\left(\ln\phi\right)^{2}}+\alpha_{0}\frac{dB}{d\ln\phi}+\beta_{0}B=\gamma_{0}\left[\frac{dK_{0}}{d\ln\phi}+K_{0}\right],\label{eq:B 2nd order}\end{equation}
where the constant coefficients $\beta_{0},\gamma_{0}$ are defined by\begin{align}
\gamma_{0} & \equiv\Lambda_{1}(v_{0})\Lambda_{3}(v_{0},R_{0})w_{\phi}^{\prime}(v_{0})\nonumber \\
 & =2\frac{c_{s}^{2}(v_{0})-w_{m}}{1+w_{m}}w_{m}^{2}R_{0}\left(1-R_{0}\right),\\
\beta_{0} & \equiv\frac{\gamma_{0}}{1-R_{0}}=2\frac{c_{s}^{2}(v_{0})-w_{m}}{1+w_{m}}w_{m}^{2}R_{0}.\end{align}

The general solution of Eq.~(\ref{eq:B 2nd order}) is the sum of an inhomogeneous
solution and the general solution of the homogeneous equation. Homogeneous solutions
are stable if both roots $\lambda_{1,2}$ of the characteristic equation\begin{equation}
\lambda^{2}+\alpha_{0}\lambda+\beta_{0}=0\end{equation}
have negative real parts,\begin{equation}
\textrm{Re}\left(\lambda_{1}\right)<0,\quad\textrm{Re}\left(\lambda_{2}\right)<0.\label{eq:condition Re lambda}\end{equation}
 This will be the case if\begin{equation}
\alpha_{0}>0,\quad\beta_{0}>0,\end{equation}
which is equivalent to the conditions\begin{equation}
\left|w_{m}\right|<1,\quad c_{s}^{2}(v_{0})>w_{m}.\label{eq:conditions cs}\end{equation}
An inhomogeneous solution of Eq.~(\ref{eq:B 2nd order}) can be expressed as\begin{align}
B(\phi) & =B_{1}(\phi)\phi^{\lambda_{1}}+B_{2}(\phi)\phi^{\lambda_{2}},\\
B_{1}(\phi) & \equiv\frac{\gamma_{0}}{\lambda_{1}-\lambda_{2}}\int^{\phi}\phi^{-\lambda_{1}-1}\left(\phi K_{0}^{\prime}+K_{0}\right)d\phi,\\
B_{2}(\phi) & \equiv\frac{\gamma_{0}}{\lambda_{2}-\lambda_{1}}\int^{\phi}\phi^{-\lambda_{2}-1}\left(\phi K_{0}^{\prime}+K_{0}\right)d\phi.\end{align}
Since the function $K_{0}(\phi)$ tends to zero at $\phi\rightarrow\infty$
by assumption, the inhomogeneous solution also tends to zero at $\phi\rightarrow\infty$
as long as the condition~(\ref{eq:condition Re lambda}) holds. This is straightforward
to show by assuming an upper bound\begin{equation}
\left|\phi K_{0}^{\prime}+K_{0}\right|<M\;\textrm{for all }\phi>\phi_{M},\end{equation}
where $\phi_{M}$ can be chosen for any $M>0$. Then the inhomogeneous solution
$B(\phi)$ is bounded for $\phi>\phi_{M}$ by\begin{equation}
\left|B(\phi)\right|<\textrm{const}\cdot M+\textrm{const}\cdot\phi^{\lambda_{1}}+\textrm{const}\cdot\phi^{\lambda_{2}},\end{equation}
which means that $B(\phi)\rightarrow0$ at $\phi\rightarrow\infty$.

Under the same assumptions, the function $A(\phi)$ will have the same behavior
at $\phi\rightarrow\infty$. We conclude that asymptotically stable solutions
$\left\{ v(\phi),R(\phi)\right\} $ approaching $\left\{ v_{0},R_{0}\right\} $
exist under the assumption $c_{s}^{2}(v_{0})\neq0$ and the further conditions~(\ref{eq:v0 determined}),
(\ref{eq:epsilon tilde nonzero}), (\ref{eq:R0 determined}), (\ref{eq:w prime nonzero}),
and~(\ref{eq:conditions cs}).%
\footnote{This is case 1 in Sec.~\ref{sec:Viable-Lagrangians-for}.%
} These conditions are similar to those derived in Ref.~\cite{Armendariz-Picon:2000ah}
under a more restrictive assumption $K(\phi)=\textrm{const}\cdot\phi^{-2}$.
Let us now investigate whether these assumptions can be relaxed further.

\subsection{Energy density $\varepsilon_{m}\propto\phi^{-2}$ and $R_{0}\neq1$, marginal
cases\label{sub:Energy-density-2}}

The last assumption used in the derivation of the stability condition~(\ref{eq:conditions cs})
was Eq.~(\ref{eq:w prime nonzero}). If $c_{s}^{2}(v_{0})=w_{m}$ while all
the other assumptions hold, we have $w_{\phi}^{\prime}(v_{0})=0$ and the equation~(\ref{eq:dB dlnphi})
for $B(\phi)$ is modified. We may then rewrite Eqs.~(\ref{eq:dA dlnphi})--(\ref{eq:dB dlnphi})
as\begin{align}
\frac{dA}{d\ln\phi} & =\left(\phi K_{0}^{\prime}+K_{0}\right)\Lambda_{1}(v_{0})-\alpha_{0}A-\frac{\Lambda_{1}(v_{0})B}{1-R_{0}},\\
\frac{dB}{d\ln\phi} & =O(A^{2}).\end{align}
Differentiating the first equation with respect to $\ln\phi$, we obtain\begin{equation}
\frac{d^{2}A}{d\left(\ln\phi\right)^{2}}=\phi\left(\phi K_{0}\right)^{\prime\prime}\Lambda_{1}(v_{0})-\alpha_{0}\frac{dA}{d\ln\phi}+O(A^{2}).\end{equation}
The second-order terms $O(A^{2})$ can be disregarded for the stability analysis.
Since the characteristic equation\begin{equation}
\lambda^{2}+\alpha_{0}\lambda=0\end{equation}
 has a zero root, the general solution $\left\{ A(\phi),B(\phi)\right\} $ will
not tend to zero at $\phi\rightarrow\infty$. Hence, no asymptotically stable
solutions exist when the condition~(\ref{eq:conditions cs}) is violated.

Another assumption, $c_{s}^{2}(v_{0})\neq0$, was used to derive Eq.~(\ref{eq:R0 determined})
that determines the allowed value of $R_{0}$. Let us briefly consider the possibility
$c_{s}^{2}(v_{0})=0$. (We note that $v\neq v_{0}$ on actual trajectories,
so stability will hold as long as the trajectories $v(\phi)$ do not reach the
regime $c_{s}^{2}(v)\leq0$.) If \begin{equation}
c_{s}^{2}(v_{0})=\frac{Q'(v_{0})}{v_{0}Q^{\prime\prime}(v_{0})}=0,\end{equation}
then $Q'(v_{0})=0$ and the asymptotic equation of state is \begin{equation}
w_{\phi}(v_{0})=\frac{Q(v_{0})}{v_{0}Q'(v_{0})-Q(v_{0})}=-1\end{equation}
as long as $Q(v_{0})\neq0$. However, we assumed a matter component with $w_{m}\neq-1$,
and so we discard the possibility that $Q(v_{0})\neq0$. If, on the other hand,
$Q(v_{0})=0$, then we must also have $\tilde{\varepsilon}_{\phi}(v_{0})=0$.
Thus $c_{s}^{2}(v_{0})\neq0$ is justified given that $\tilde{\varepsilon}_{\phi}(v_{0})\neq0$.

Relaxing the assumption $\tilde{\varepsilon}_{\phi}(v_{0})\neq0$ requires some
more work. If $\tilde{\varepsilon}_{\phi}(v_{0})=0$, then we cannot conclude
that $K(\phi)\propto\phi^{-2}$ at $\phi\rightarrow\infty$; the function $K(\phi)$
remains undetermined even though we know that $\varepsilon_{\phi}(\phi)=K(\phi)\tilde{\varepsilon}_{\phi}(v)\propto\phi^{-2}$.
The analysis after Eq.~(\ref{eq:v0 determined}) needs to be modified as follows.
The finiteness of $w_{\phi}$,\begin{equation}
w_{\phi}(v_{0})=\lim_{v\rightarrow v_{0}}\frac{Q(v)}{\tilde{\varepsilon}_{\phi}(v)}<\infty,\end{equation}
requires that $Q(v_{0})=0$ and thus (since $v_{0}\neq0$) also $Q'(v_{0})=0$.
In general, we may suppose that $Q(v)$ has an expansion \begin{equation}
Q(v)=\frac{Q_{0}}{nv_{0}}\left(v-v_{0}\right)^{n}\left[1+O(v-v_{0})\right],\label{eq:expansion Q}\end{equation}
where $Q_{0}$ is a nonzero constant and $n\geq2$. In this case we have the
expansions\begin{align}
\tilde{\varepsilon}_{\phi}(v) & =Q_{0}\left(v-v_{0}\right)^{n-1}\left[1+O(v-v_{0})\right],\label{eq:expansion eps tilde}\\
w_{\phi}(v) & =\frac{v-v_{0}}{nv_{0}}\left[1+O(v-v_{0})\right],\label{eq:expansion w k}\\
c_{s}^{2}(v) & =\frac{v-v_{0}}{\left(n-1\right)v_{0}}\left[1+O(v-v_{0})\right].\label{eq:expansion c s}\end{align}
It follows that $w_{\phi}(v_{0})=0$, so the only possibility for tracking is
$w_{m}=0$. Also, the only admissible solutions are those with $v(\phi)>v_{0}$,
meaning that $A(\phi)<0$ and $Q_{0}>0$. Let us now perform a stability analysis
of these solutions. Substituting the ansatz~(\ref{eq:v R ansatz}) into Eqs.~(\ref{eq:v equ clever})--(\ref{eq:R EOM})
and keeping only the leading terms in the perturbation variables $A(\phi)$
and $B(\phi)$, we obtain \begin{align}
\frac{dA}{d\phi} & =\frac{\left(-A\right)}{n-1}\frac{K'}{K}+\frac{3\kappa}{\left(n-1\right)v_{0}}\sqrt{\frac{K(\phi)Q_{0}}{1-R_{0}}}\left(-A\right)^{\left(n+1\right)/2},\label{eq:equation A marginal 1}\\
\frac{dB}{d\phi} & =-\frac{3\kappa}{nv_{0}^{2}}R_{0}\sqrt{1-R_{0}}\sqrt{K(\phi)Q_{0}}\left(-A\right)^{\left(n+1\right)/2}.\end{align}
For the purposes of a stability analysis, it is sufficient to note that Eq.~(\ref{eq:equation A marginal 1})
does not involve $B(\phi)$. One can solve Eq.~(\ref{eq:equation A marginal 1})
explicitly for $A(\phi)$ and find such $K(\phi)$ that the general solution
for $A(\phi)$ tends to zero at $\phi\rightarrow\infty$; for instance, $K(\phi)\propto\phi^{r}$
with $r>-2$. However, the general solution for $B(\phi)$ is\begin{equation}
B(\phi)=B_{0}-\textrm{const}\cdot\int_{\phi_{0}}^{\phi}\sqrt{K(\phi)}\left(-A\right)^{\left(n+1\right)/2}d\phi,\end{equation}
 where $B_{0}$ is an arbitrary integration constant. It follows that $B(\phi)$
will either diverge or tend to an arbitrary constant of integration at $\phi\rightarrow\infty$.
Hence, the general perturbation will not tend to zero at large $\phi$. We conclude
that no asymptotically stable solutions exist when $\tilde{\varepsilon}_{\phi}(v_{0})=0$.

\subsection{Energy density $\varepsilon_{m}\propto\phi^{-2}$ and $R_{0}=1$, main case
\label{sub:Energy-density-dominates-1}}

We use the ansatz $R(\phi)=1-B(\phi)$, where the function $B(\phi)$ is positive
and tends to zero monotonically as $\phi\rightarrow\infty$. Since $dR/d\phi>0$,
it follows from Eq.~(\ref{eq:R equ 0}) that $w_{m}<w_{\phi}(v(\phi))$ for
all sufficiently large $\phi$. Thus, any asymptotically stable solutions will
necessarily satisfy the condition \begin{equation}
w_{m}\leq w_{\phi}(v_{0}).\end{equation}

Since $R\rightarrow1$ as $\phi\rightarrow\infty$, we have $\varepsilon_{\text{tot}}(\phi)\propto\varepsilon_{m}(\phi)\propto\phi^{-2}$,
so we may write \begin{equation}
\varepsilon_{\text{tot}}(\phi)\approx E_{0}\phi^{-2},\quad\phi\rightarrow\infty,\label{eq:e tot asympt 1}\end{equation}
where $E_{0}$ is a nonzero constant. The value of $E_{0}$ can be related to
other parameters by using Eq.~(\ref{eq:epsilon m EOM}), rewritten as\begin{equation}
\frac{d\ln\varepsilon_{m}}{d\ln\phi}=-\frac{3\kappa}{v}\sqrt{\phi^{2}\varepsilon_{m}}\left(1+w_{m}\right),\end{equation}
which yields, in the limit $\phi\rightarrow\infty$,\begin{equation}
2=\frac{3\kappa}{v_{0}}\sqrt{E_{0}}\left(1+w_{m}\right).\label{eq:E0 relation}\end{equation}
 Expressing $\varepsilon_{\text{tot}}$ through $\varepsilon_{\phi}$, we have\begin{equation}
E_{0}\phi^{-2}\approx\varepsilon_{\text{tot}}=\frac{\varepsilon_{\phi}}{1-R}=\frac{\tilde{\varepsilon}_{\phi}(v)K(\phi)}{B};\end{equation}
hence\begin{equation}
B(\phi)\approx\frac{\tilde{\varepsilon}_{\phi}(v)\phi^{2}K(\phi)}{E_{0}},\quad\phi\rightarrow\infty.\label{eq:B asymptotic condition}\end{equation}

We now assume that $\tilde{\varepsilon}_{\phi}(v_{0})\neq0$; the case $\tilde{\varepsilon}_{\phi}(v_{0})=0$
will be considered later. If $\tilde{\varepsilon}_{\phi}(v_{0})\neq0$, it follows
that\begin{equation}
B(\phi)\approx\frac{\tilde{\varepsilon}_{\phi}(v_{0})}{E_{0}}\phi^{2}K(\phi),\quad\phi\rightarrow\infty.\label{eq:B asympt 1}\end{equation}
Rewriting Eq.~(\ref{eq:R equ 0}) as\begin{equation}
\frac{d\ln\left(1-R\right)}{d\ln\phi}=\frac{3\kappa R}{v}\sqrt{\phi^{2}\varepsilon_{\text{tot}}}\left(w_{m}-w_{\phi}(v)\right)\label{eq:ln 1-R equ}\end{equation}
and substituting Eqs.~(\ref{eq:e tot asympt 1}) and (\ref{eq:B asympt 1}),
we find for large $\phi$\begin{equation}
\frac{d\ln\left(\phi^{2}K(\phi)\right)}{d\ln\phi}\approx\frac{3\kappa\sqrt{E_{0}}}{v_{0}}\left(w_{m}-w_{\phi}(v)\right)=2\frac{w_{m}-w_{\phi}(v)}{1+w_{m}}.\label{eq:K equ 1}\end{equation}
It is now clear that the possible asymptotic behavior of $K(\phi)$ at $\phi\rightarrow\infty$
depends on whether $w_{\phi}(v)$ tends to $w_{m}$ at large $\phi$, i.e.~on
whether or not $w_{\phi}(v_{0})=w_{m}$. (We note that the value of $v_{0}$
is yet to be determined by the analysis that follows.)

Considering the interesting case $w_{\phi}(v_{0})\neq w_{m}$, we find that
the right-hand side of Eq.~(\ref{eq:K equ 1}) tends to a negative constant
as $\phi\rightarrow\infty$. Denoting that constant by $-\mu$, where \begin{equation}
\mu\equiv2\frac{w_{\phi}(v_{0})-w_{m}}{1+w_{m}}>0,\label{eq:mu def}\end{equation}
 and integrating Eq.~(\ref{eq:K equ 1}), we infer the following asymptotic
behavior of $K(\phi)$,\begin{equation}
K(\phi)\propto\phi^{-2-\mu}K_{0}(\phi),\quad\phi\rightarrow\infty,\end{equation}
where $K_{0}(\phi)$ is an auxiliary function that satisfies\begin{equation}
\lim_{\phi\rightarrow\infty}\frac{d\ln K_{0}(\phi)}{d\ln\phi}=0.\end{equation}
This condition is equivalent to\begin{equation}
\lim_{\phi\rightarrow\infty}\frac{\ln K_{0}(\phi)}{\ln\phi}=0.\end{equation}
Thus, the function $K_{0}(\phi)$ may go to a constant at large $\phi$, or
may grow or decay slower than any power of $\phi$; examples of admissible functions
$K_{0}(\phi)$ are\begin{equation}
K_{0}(\phi)=\left(\ln\phi\right)^{p};\quad K_{0}(\phi)=\exp\left(C_{1}\left(\ln\phi\right)^{s}\right),\quad\left|s\right|<1.\end{equation}
 With any such $K_{0}(\phi)$, solutions of the currently considered type are
possible only for Lagrangians $p=K(\phi)Q(v)$ with\begin{equation}
K(\phi)=\phi^{-2\alpha}K_{0}(\phi),\label{eq:K ansatz 3}\end{equation}
 where\begin{equation}
\alpha\equiv\frac{2+\mu}{2}=\frac{1+w_{\phi}(v_{0})}{1+w_{m}}>1.\label{eq:K asympt 2 and w}\end{equation}
 For a given Lagrangian of this type, the possible values of $v_{0}$ are fixed
by Eq.~(\ref{eq:K asympt 2 and w}). If Eq.~(\ref{eq:K asympt 2 and w}) is
not satisfied for any such $v_{0}$, solutions of this type do not exist. The
value $w_{\phi}(v_{0})$ is determined by Eq.~(\ref{eq:K asympt 2 and w})
as\begin{equation}
w_{\phi}(v_{0})=\left(1+w_{m}\right)\alpha-1.\label{eq:wk v0 fixed by K asympt 2 and w}\end{equation}
Since $w_{m}\neq-1$, we must have $w_{\phi}(v_{0})\neq-1$ also.

It remains to investigate the asymptotic stability of the general solution.
Since $B(\phi)$ must satisfy Eq.~(\ref{eq:B asymptotic condition}), we may
write an ansatz\begin{equation}
B(\phi)=\frac{\tilde{\varepsilon}_{\phi}(v_{0})}{E_{0}}\phi^{2}K(\phi)\left(1+C(\phi)\right),\label{eq:B ansatz 3}\end{equation}
where $C(\phi)$ is a new perturbation variable. Hence, we substitute Eq.~(\ref{eq:K ansatz 3})
together with the ansatz \begin{align}
v(\phi) & =v_{0}-A(\phi),\\
R(\phi) & =1-\frac{\tilde{\varepsilon}_{\phi}(v_{0})}{E_{0}}\phi^{2}K(\phi)\left(1+C(\phi)\right),\\
\varepsilon_{\text{tot}}(v,\phi) & =E_{0}\phi^{-2}\frac{\tilde{\varepsilon}_{\phi}(v)}{\tilde{\varepsilon}_{\phi}(v_{0})}\frac{1}{1+C(\phi)},\end{align}
 into Eqs.~(\ref{eq:v equ clever}) and (\ref{eq:ln 1-R equ}). Using Eqs.~(\ref{eq:E0 relation}),
(\ref{eq:mu def}), and (\ref{eq:K ansatz 3}), we obtain at an intermediate
step the equations\begin{align}
\frac{dA}{d\phi}= & \:\left[-\frac{2\alpha}{\phi}+\left(\ln K_{0}\right)^{\prime}\right]\Lambda_{1}(v)\nonumber \\
 & \:+\frac{\phi^{-1}}{\sqrt{1+C}}\frac{2v_{0}}{1+w_{m}}\frac{\Lambda_{2}(v)}{3\kappa\sqrt{\tilde{\varepsilon}_{\phi}(v_{0})}},\label{eq:A equ 4}\\
\frac{1}{1+C}\frac{dC}{d\phi}= & \:\mu\phi^{-1}-\left(\ln K_{0}\right)^{\prime}\nonumber \\
 & -\frac{\mu\phi^{-1}}{\sqrt{1+C}}\frac{\Lambda_{4}(v)}{\Lambda_{4}(v_{0})}\left[1+O(\phi^{-\mu})\right],\label{eq:C equ 4}\end{align}
where the functions $\Lambda_{1}(v)$ and $\Lambda_{2}(v)$ were defined by
Eqs.~(\ref{eq:Lambda1 def})--(\ref{eq:Lambda2 def}), while the new auxiliary
function $\Lambda_{4}(v)$ is defined by\begin{equation}
\Lambda_{4}(v)\equiv3\kappa\sqrt{\tilde{\varepsilon}_{\phi}(v)}\frac{w_{\phi}(v)-w_{m}}{v}.\label{eq:Lambda4 def}\end{equation}
 In the present case, the identity\begin{equation}
2\alpha\Lambda_{1}(v_{0})=\frac{2v_{0}}{1+w_{m}}\frac{\Lambda_{2}(v_{0})}{3\kappa\sqrt{\tilde{\varepsilon}_{\phi}(v_{0})}}\end{equation}
 holds due to Eq.~(\ref{eq:K asympt 2 and w}). 

We now linearize Eqs.~(\ref{eq:A equ 4})--(\ref{eq:C equ 4}) with respect
to the perturbation variables $A$ and $C$. To simplify the linearized equations,
we use Eqs.~(\ref{eq:derivative of wk}), (\ref{eq:E0 relation}), and the
definition~(\ref{eq:K asympt 2 and w}) of $\alpha$. (We note that $\Lambda_{1}(v_{0})\neq0$;
otherwise, we would have $c_{s}^{2}(v_{0})=0$, which contradicts the earlier
assumptions $\tilde{\varepsilon}_{\phi}(v_{0})\neq0$ and $w_{\phi}(v_{0})\neq-1$.)
After some algebra, we find (to the leading order)\begin{align}
\frac{dA}{d\ln\phi} & =2\alpha\Lambda_{2}(v_{0})\left(\frac{\Lambda_{1}(v)}{\Lambda_{2}(v)}\right)_{v_{0}}^{\prime}\negmedspace A\nonumber \\
 & \quad+\Lambda_{1}(v_{0})\left[\frac{d\ln K_{0}}{d\ln\phi}-\alpha C\right],\label{eq:A equ 5}\\
\frac{dC}{d\ln\phi} & =-\frac{d\ln K_{0}}{d\ln\phi}+\frac{1}{2}\mu C+\mu\frac{\Lambda_{4}^{\prime}(v_{0})}{\Lambda_{4}(v_{0})}A+\mu\frac{\tilde{\varepsilon}_{\phi}(v_{0})}{E_{0}}\phi^{-\mu}.\label{eq:C equ 5}\end{align}
 This is an inhomogeneous linear system for $A(\phi)$ and $C(\phi)$. The analysis
of the asymptotic stability is similar to that after Eq.~(\ref{eq:condition Re lambda}).
Since all the inhomogeneous terms are decaying at $\phi\rightarrow\infty$,
it suffices to require that both the eigenvalues of the homogeneous system have
negative real parts. For a homogeneous system of the form\begin{align}
\frac{dA}{d\ln\phi} & =\beta_{1}A+\beta_{2}C,\label{eq:A hom equ}\\
\frac{dC}{d\ln\phi} & =\gamma_{1}A+\gamma_{2}C,\label{eq:C hom equ}\end{align}
the characteristic equation is\begin{equation}
\lambda^{2}-\left(\beta_{1}+\gamma_{2}\right)\lambda+\left(\beta_{1}\gamma_{2}-\beta_{2}\gamma_{1}\right)=0,\end{equation}
and the stability conditions are\begin{equation}
\beta_{1}+\gamma_{2}<0,\quad\beta_{1}\gamma_{2}-\beta_{2}\gamma_{1}>0.\label{eq:stability conditions beta gamma}\end{equation}
Presently, the constants $\beta_{1},\beta_{2},\gamma_{1},\gamma_{2}$ can be
read off from Eqs.~(\ref{eq:A equ 5})--(\ref{eq:C equ 5}); simplifying, we
obtain\begin{align}
\beta_{1} & =-\alpha\frac{1-w_{\phi}(v_{0})}{1+w_{\phi}(v_{0})},\quad\beta_{2}=-\alpha\frac{c_{s}^{2}(v_{0})v_{0}}{1+w_{\phi}(v_{0})},\\
\gamma_{1} & =\frac{\mu}{v_{0}c_{s}^{2}(v_{0})}\biggl[\frac{\left(c_{s}^{2}(v_{0})-w_{\phi}(v_{0})\right)\left(1+w_{m}\right)}{w_{\phi}(v_{0})-w_{m}}\nonumber \\
 & \quad+\frac{1-w_{\phi}(v_{0})}{2}\biggr],\quad\gamma_{2}=\frac{\mu}{2}.\end{align}
 The stability conditions~(\ref{eq:stability conditions beta gamma}) can be
simplified to\begin{equation}
\frac{1-w_{\phi}(v_{0})}{1+w_{\phi}(v_{0})}>\frac{\mu}{2\alpha},\quad\frac{c_{s}^{2}(v_{0})-w_{\phi}(v_{0})}{w_{\phi}(v_{0})-w_{m}}>0.\end{equation}
Since $w_{\phi}(v_{0})>w_{m}$ for solutions of the present type, while $2\alpha=\mu+2$,
the stability conditions (together with the condition $w_{\phi}(v_{0})>w_{m}$)
are\begin{equation}
w_{m}<w_{\phi}(v_{0})<\frac{1}{1+\mu},\quad c_{s}^{2}(v_{0})>w_{\phi}(v_{0}).\end{equation}
Using Eq.~(\ref{eq:wk v0 fixed by K asympt 2 and w}), we can transform the
first of these conditions into a condition for $\alpha$:\begin{equation}
1<\alpha<\frac{1}{2}+\frac{1}{1+w_{m}},\quad c_{s}^{2}(v_{0})>w_{\phi}(v_{0}).\end{equation}
The first inequality above will define a nonempty interval of $\alpha$ only
if $\left|w_{m}\right|<1$. These are the final conditions for the asymptotic
stability of the solutions obtained under the assumptions $\tilde{\varepsilon}_{\phi}(v_{0})\neq0$,
$w_{\phi}(v_{0})\neq w_{m}$, and (\ref{eq:K asympt 2 and w}).%
\footnote{This is case 3 in Sec.~\ref{sec:Viable-Lagrangians-for}.%
}

\subsection{Energy density $\varepsilon_{m}\propto\phi^{-2}$ and $R_{0}=1$, marginal
cases}

The analysis in the previous section used the assumptions $\tilde{\varepsilon}_{\phi}(v_{0})\neq0$
and $w_{\phi}(v_{0})\neq w_{m}$. In this section we lift these assumption,
in the reverse order used. 

If $w_{\phi}(v_{0})=w_{m}$ while $\tilde{\varepsilon}_{\phi}(v_{0})\neq0$,
then we may continue the arguments starting with Eq.~(\ref{eq:K equ 1}). Note
that Eqs.~(\ref{eq:E0 relation}) and (\ref{eq:B asympt 1}) still hold. Since
the right-hand side of Eq.~(\ref{eq:K equ 1}) tends to zero at $\phi\rightarrow\infty$,
it follows that\begin{equation}
\lim_{\phi\rightarrow\infty}\frac{d\ln\left(\phi^{2}K(\phi)\right)}{d\ln\phi}=0.\end{equation}
This condition is equivalent to\begin{equation}
\lim_{\phi\rightarrow\infty}\frac{\ln K(\phi)}{\ln\phi}=-2.\label{eq:K phi limit 0}\end{equation}
Also, according to Eq.~(\ref{eq:B asympt 1}) we can have $B(\phi)\rightarrow0$
only if \begin{equation}
\lim_{\phi\rightarrow\infty}\phi^{2}K(\phi)=0.\label{eq:K phi limit 01}\end{equation}
So the function $K(\phi)$ cannot have a power-law asymptotic other than $\phi^{-2}$;
more precisely, for any $\varepsilon>0$ and for large enough $\phi$ we must
have\begin{equation}
K(\phi)<\phi^{-2+\varepsilon},\quad K(\phi)>\phi^{-2-\varepsilon},\quad\phi\rightarrow\infty.\end{equation}
However, a non-power law asymptotic behavior at $\phi\rightarrow\infty$ is
still admissible, for instance $K(\phi)\propto\phi^{-2}\left(\ln\phi\right)^{-s}$,
where $s>0$ to allow $B(\phi)\rightarrow0$ according to Eq.~(\ref{eq:B asympt 1}).
Rather than assume a particular form of $K(\phi)$, we will perform the analysis
for arbitrary $K(\phi)$ satisfying Eq.~(\ref{eq:K phi limit 0}). 

We again use the ansatz~(\ref{eq:B ansatz 3}) to linearize Eqs.~(\ref{eq:v equ clever})
and (\ref{eq:ln 1-R equ}). After some algebra, we find (to the leading order)\begin{align}
\frac{dA}{d\ln\phi} & =\left[-2+\frac{d\ln\left(\phi^{2}K\right)}{d\ln\phi}\right]\Lambda_{1}(v)+\frac{\Lambda_{2}(v)}{\sqrt{1+C}}\sqrt{\frac{E_{0}}{\tilde{\varepsilon}_{\phi}(v_{0})}},\label{eq:A equ 6}\\
\frac{dC}{d\ln\phi} & =-\frac{d\ln\left(\phi^{2}K\right)}{d\ln\phi}-\frac{\Lambda_{4}(v)}{\sqrt{1+C}}\sqrt{\frac{E_{0}}{\tilde{\varepsilon}_{\phi}(v_{0})}},\label{eq:C equ 6}\end{align}
where the auxiliary functions $\Lambda_{1}(v)$, $\Lambda_{2}(v)$, and $\Lambda_{4}(v)$
were defined above by Eqs.~(\ref{eq:Lambda1 def}), (\ref{eq:Lambda2 def}),
and (\ref{eq:Lambda4 def}). Since $w_{\phi}(v_{0})=w_{m}$ and $c_{s}^{2}(v_{0})\neq0$,
we have the relationship,\begin{equation}
2\Lambda_{1}(v_{0})=\Lambda_{2}(v_{0})\sqrt{\frac{E_{0}}{\tilde{\varepsilon}_{\phi}(v_{0})}}\neq0,\end{equation}
and then, assuming for the moment that $\Lambda_{4}^{\prime}(v_{0})\neq0$,
we can linearize Eqs.~(\ref{eq:A equ 6})--(\ref{eq:C equ 6}) as\begin{align}
\frac{dA}{d\ln\phi} & =2\Lambda_{2}(v_{0})\left(\frac{\Lambda_{1}(v)}{\Lambda_{2}(v)}\right)_{v_{0}}^{\prime}A-C\Lambda_{1}(v_{0})\nonumber \\
 & \quad+\frac{d\ln\left(\phi^{2}K\right)}{d\ln\phi}\Lambda_{1}(v_{0}),\label{eq:A equ 7}\\
\frac{dC}{d\ln\phi} & =\frac{2v_{0}}{1+w_{m}}\frac{\Lambda_{4}^{\prime}(v_{0})}{3\kappa\sqrt{\tilde{\varepsilon}_{\phi}(v_{0})}}A-\frac{d\ln\left(\phi^{2}K\right)}{d\ln\phi}.\label{eq:C equ 7}\end{align}
The stability analysis proceeds as before, since all the inhomogeneous terms
are decaying at $\phi\rightarrow\infty$. The resulting conditions are simplified
to\begin{equation}
\frac{1-w_{m}}{1+w_{m}}>0,\quad\frac{c_{s}^{2}-w_{m}}{1+w_{m}}>0,\end{equation}
and further to\begin{equation}
\left|w_{m}\right|<1,\quad c_{s}^{2}>w_{m}.\end{equation}
These conditions are the same as the standard stability conditions for a tracker
solution. Under these conditions, a tracker solution with $w_{\phi}(v_{0})=w_{m}$
exists as long as $K(\phi)$ satisfies Eqs.~(\ref{eq:K phi limit 0})--(\ref{eq:K phi limit 01}).%
\footnote{This is case 4 in Sec.~\ref{sec:Viable-Lagrangians-for}.%
}

Finally, we analyze the case $\tilde{\varepsilon}_{\phi}(v_{0})=0$. Since Eqs.~(\ref{eq:e tot asympt 1}),
(\ref{eq:E0 relation}), and (\ref{eq:B asymptotic condition}) still hold for
an asymptotically stable solution, we are motivated to use the ansatz\begin{align}
v(\phi) & =v_{0}-A(\phi),\label{eq:B ansatz 4}\\
R(\phi) & =1-B(v,\phi),\\
B(v,\phi) & \equiv\frac{\tilde{\varepsilon}_{\phi}(v)}{E_{0}}\phi^{2}K(\phi)\left(1+C(\phi)\right).\end{align}
We first derive the \emph{exact} equations of motion for the variables $A(\phi),C(\phi)$
from Eqs.~(\ref{eq:v equ clever}) and (\ref{eq:ln 1-R equ}):\begin{align}
\frac{dA}{d\ln\phi} & =\frac{vc_{s}^{2}(v)}{1+w_{\phi}(v)}\frac{d\ln K}{d\ln\phi}+\frac{2v_{0}}{1+w_{m}}\frac{c_{s}^{2}(v)}{\sqrt{1+C}},\label{eq:A equ very clever}\\
\frac{d\ln\left(1+C\right)}{d\ln\phi} & =\frac{2}{v}A+\frac{2v_{0}}{v}\left(\frac{1}{\sqrt{1+C}}-1\right)\nonumber \\
 & \quad-\frac{2v_{0}}{v}\frac{B(v,\phi)}{\sqrt{1+C(\phi)}}\frac{w_{m}-w_{\phi}(v)}{1+w_{m}}.\label{eq:1C equ very clever}\end{align}
 Then the stability analysis consists of checking that the general solution
involves functions $A(\phi),B(v,\phi),C(\phi)$ that decay as $\phi\rightarrow\infty$.
Since $v_{0}\neq0$, the expansions~(\ref{eq:expansion eps tilde})--(\ref{eq:expansion c s})
hold with $n\geq2$; we note that $A<0$ to guarantee $c_{s}^{2}>0$, and that
$w_{\phi}(v_{0})=0$. The leading-order terms in Eq.~(\ref{eq:A equ very clever})
are\begin{equation}
\frac{d\left|A\right|}{d\ln\phi}=-\frac{\left|A\right|}{n-1}\left(\frac{d\ln K(\phi)}{d\ln\phi}+\frac{2}{1+w_{m}}\right),\end{equation}
and the general solution is\begin{equation}
\left|A\right|=A_{0}\left[\phi^{\frac{2}{1+w_{m}}}K(\phi)\right]^{-\frac{1}{n-1}}.\end{equation}
Since $n\geq2$, solutions $A(\phi)$ decay at $\phi\rightarrow\infty$ as long
as \begin{equation}
K(\phi)\phi^{\frac{2}{1+w_{m}}}\rightarrow\infty,\quad\phi\rightarrow\infty.\end{equation}
The function $B(v,\phi)$ is then expressed as\begin{equation}
B(v,\phi)=\frac{Q_{0}A_{0}^{n-1}}{E_{0}}\phi^{\frac{2w_{m}}{1+w_{m}}}\left(1+C(\phi)\right),\end{equation}
and its decay at $\phi\rightarrow\infty$ requires that $-1<w_{m}<0$. The leading
terms of Eq.~(\ref{eq:1C equ very clever}) are\begin{align*}
\frac{dC}{d\ln\phi} & =\frac{2}{v_{0}}A-C-\left(1-\frac{1}{2}C\right)B(v,\phi)\frac{w_{m}}{1+w_{m}}.\end{align*}
Since the homogeneous solution $C(\phi)\propto\phi^{-1}$ decays as a power
of $\phi$, while the inhomogeneous terms all decay at $\phi\rightarrow\infty$,
the general solution $C(\phi)$ will also decay at $\phi\rightarrow\infty$.
Thus, we find a family of asymptotically stable solutions corresponding to a
value $v_{0}$ such that Eq.~(\ref{eq:expansion Q}) holds, in case $w_{m}<0$
and for Lagrangians with $K(\phi)$ that either does not decay at large $\phi$,
or decays slower than $\phi^{-2/(1+w_{m})}$.%
\footnote{This is case 5 in Sec.~\ref{sec:Viable-Lagrangians-for}.%
}

\subsection{Domination by $k$-essence, $v_{0}\neq0$, main case}

We now consider the case $R_{0}=0$. In this case, the matter component becomes
subdominant at late times, so $\varepsilon_{\text{tot}}\approx\varepsilon_{\phi}$
at $\phi\rightarrow\infty$. According to Eq.~(\ref{eq:epsilon EOM}), we have
at late times\begin{align}
\frac{d}{d\phi}\varepsilon_{\phi}(\phi) & =-\frac{3\kappa}{v}\sqrt{\varepsilon_{\text{tot}}}\left(1+w_{\phi}\right)\varepsilon_{\phi}\nonumber \\
 & \approx-\frac{3\kappa}{v_{0}}\varepsilon_{\phi}^{3/2}\left(1+w_{\phi}(v)\right),\label{epsilon k 3}\end{align}
thus the asymptotic behavior of $\varepsilon_{\phi}(\phi)$ depends on whether
or not $w_{\phi}(v_{0})=-1$. With $v_{0}\neq0$, one can have $w_{\phi}(v_{0})=-1$
only if $Q'(v_{0})=0$, which entails \begin{equation}
c_{s}^{2}(v_{0})=\frac{1}{v_{0}}\lim_{v\rightarrow v_{0}}\frac{Q'(v)}{Q''(v)}=\frac{1}{v_{0}}\lim_{v\rightarrow v_{0}}\frac{1}{\left(\ln Q'(v)\right)^{\prime}}=0.\end{equation}

Let us postpone the consideration of the case $c_{s}(v_{0})=0$; thus, presently
we have $w_{\phi}(v_{0})\neq-1$. In that case, the asymptotic behavior of $\varepsilon_{\phi}(\phi)$
and $\varepsilon_{\text{tot}}(\phi)$ can be expressed as\begin{equation}
\varepsilon_{\text{tot}}(\phi)\approx\varepsilon_{\phi}(\phi)\approx E_{0}\phi^{-2},\label{eq:epsilon k asympt 44}\end{equation}
where the constant $E_{0}$ is given by\begin{equation}
3\kappa\sqrt{E_{0}}=\frac{2v_{0}}{1+w_{\phi}(v_{0})},\label{eq:E0 cond 4}\end{equation}
due to Eq.~(\ref{epsilon k 3}). We use Eqs.~(\ref{eq:v equ clever})--(\ref{eq:R EOM})
to describe asymptotically stable solutions. Since on such solutions $R(\phi)$
approaches zero while remaining positive, we must have $w_{\phi}(v)<w_{m}$
at late times. Computing the limit of Eq.~(\ref{eq:v equ clever}) as $\phi\rightarrow\infty$
and using Eq.~(\ref{eq:E0 cond 4}), we find\begin{align}
0 & =\lim_{\phi\rightarrow\infty}\frac{dv}{d\ln\phi}=\lim_{\phi\rightarrow\infty}\phi c_{s}^{2}(v)\left[\frac{\left(\ln K\right)_{,\phi}v}{1+w_{\phi}(v)}+3\kappa\sqrt{\varepsilon_{\text{tot}}}\right]\nonumber \\
 & =\frac{v_{0}}{1+w_{\phi}(v_{0})}\lim_{\phi\rightarrow\infty}c_{s}^{2}(v)\left[\frac{d\ln K(\phi)}{d\ln\phi}+2\right].\label{eq:A equ 8}\end{align}
The right-hand side of Eq.~(\ref{eq:A equ 8}) can vanish at $\phi\rightarrow\infty$
if, for instance, $c_{s}^{2}(v_{0})=0$. We postpone the consideration of the
case $c_{s}^{2}(v_{0})=0$ and presently assume that $c_{s}^{2}(v_{0})\neq0$,
which (together with $v_{0}\neq0$) also implies $\tilde{\varepsilon}_{\phi}(v_{0})\neq0$.
Then $\varepsilon_{\phi}(\phi)\propto\phi^{-2}$ entails $K(\phi)\propto\phi^{-2}$
at $\phi\rightarrow\infty$; accordingly, the right-hand side of Eq.~(\ref{eq:A equ 8})
vanishes at $\phi\rightarrow\infty$ due to\begin{equation}
\lim_{\phi\rightarrow\infty}\frac{d\ln K}{d\ln\phi}=-2.\end{equation}
 By absorbing a constant into $Q(v)$ if necessary, we may express $K(\phi)$
as \begin{equation}
K(\phi)=\frac{1+K_{0}(\phi)}{\phi^{2}},\quad\lim_{\phi\rightarrow\infty}K_{0}(\phi)=0.\end{equation}
 This is the familiar form of the function $K(\phi)$, shown by Eq.~(\ref{eq:K ansatz 2})
in Sec.~\ref{sub:Energy-density-1}.

For these $K(\phi)$, the condition~(\ref{eq:E0 cond 4}) becomes \begin{equation}
3\kappa\sqrt{E_{0}}=3\kappa\sqrt{\tilde{\varepsilon}_{\phi}(v_{0})}=\frac{2v_{0}}{1+w_{\phi}(v_{0})},\label{eq:v0 equ 4}\end{equation}
which is an equation for determining the admissible values of $v_{0}$. For
these $v_{0}$, we linearize Eqs.~(\ref{eq:v equ clever})--(\ref{eq:R EOM})
using the ansatz\begin{equation}
v=v_{0}-A(\phi),\quad R=B(\phi),\label{eq:v0 R 4 ansatz}\end{equation}
where $A(\phi),B(\phi)$ tend to zero as $\phi\rightarrow\infty$. The manipulations
with Eq.~(\ref{eq:v equ clever}) are the same as those in Sec.~\ref{sub:Energy-density-1};
the result of the linearization is quite similar to Eq.~(\ref{eq:dA dlnphi})
with $R_{0}=0$ and without the relationship $w_{\phi}(v_{0})=w_{m}$,\begin{equation}
\frac{dA}{d\ln\phi}=\left(\phi K_{0}^{\prime}+K_{0}\right)\Lambda_{1}(v_{0})-\frac{1-w_{\phi}(v_{0})}{1+w_{\phi}(v_{0})}A-\Lambda_{1}(v_{0})B.\end{equation}
The linearized form of Eq.~(\ref{eq:R EOM}) is\begin{align}
\frac{dB}{d\ln\phi} & =-\frac{3\kappa\sqrt{E_{0}}}{v_{0}}\left(w_{m}-w_{\phi}(v_{0})\right)B\nonumber \\
 & =-2\frac{w_{m}-w_{\phi}(v_{0})}{1+w_{\phi}(v_{0})}B.\end{align}
Since the equation for $B(\phi)$ does not involve $A(\phi)$, and since $w_{m}>w_{\phi}(v_{0})$,
all solutions $B(\phi)$ decay, and thus all solutions $A(\phi)$ also decay
as long as \begin{equation}
\frac{1-w_{\phi}(v_{0})}{1+w_{\phi}(v_{0})}>0,\end{equation}
which is equivalent to $\left|w_{\phi}(v_{0})\right|<1$. Therefore, solutions
are asymptotically stable under the conditions~(\ref{eq:v0 equ 4}), $\left|w_{\phi}(v_{0})\right|<1$,
$w_{m}>w_{\phi}(v_{0})$, and $c_{s}(v_{0})\neq0$.%
\footnote{This is case 2 in Sec.~\ref{sec:Viable-Lagrangians-for}.%
}

\subsection{Domination by $k$-essence, $v_{0}\neq0$, marginal cases}

In this section we continue considering the case $R_{0}=0$, $v_{0}\neq0$,
and examine the possibility that $c_{s}^{2}(v_{0})=0$. In that case, we have
$Q'(v_{0})=0$ as well, which fixes admissible values of $v_{0}$. There are
two further possibilities: either $Q(v_{0})\neq0$ or $Q(v_{0})=0$.

If $Q(v_{0})\equiv Q_{0}\neq0$, then $Q(v)$ can be expanded about $v=v_{0}$
as\begin{equation}
Q(v)=Q_{0}+Q_{1}\left(v-v_{0}\right)^{n}\left[1+O(v-v_{0})\right],\label{eq:Q expansion Q0}\end{equation}
where $n\geq2$. One readily obtains the expansions\begin{align}
\tilde{\varepsilon}_{\phi}(v) & =-Q_{0}+nv_{0}Q_{1}\left(v-v_{0}\right)^{n-1}\left[1+O(v-v_{0})\right],\\
w_{\phi}(v) & =-1+\frac{nv_{0}Q_{1}}{-Q_{0}}\left(v-v_{0}\right)^{n-1}\left[1+O(v-v_{0})\right],\label{eq:wk expansion Q0}\\
c_{s}^{2}(v) & =\frac{1}{v_{0}}\frac{v-v_{0}}{n-1}\left[1+O(v-v_{0})\right].\label{eq:cs expansion Q0}\end{align}
It is clear that one must have $K(\phi)Q_{0}<0$ due to the positivity of the
energy density. For convenience, let us assume that $K(\phi)>0$ and $Q_{0}<0$.
Substituting the expansions above into Eqs.~(\ref{eq:v equ clever})--(\ref{eq:R EOM})
together with the ansatz~(\ref{eq:v0 R 4 ansatz}) and neglecting the subleading
terms, we obtain \begin{align}
\frac{dA}{d\phi} & =\frac{1}{v_{0}}\frac{-A}{n-1}\left[\frac{\left|Q_{0}\right|\left(\ln K\right)_{,\phi}}{nQ_{1}\left(-A\right)^{n-1}}+3\kappa\sqrt{\left|Q_{0}\right|K(\phi)}\right],\label{eq:A equ 9}\\
\frac{dB}{d\phi} & =-\frac{3\kappa}{v_{0}}B\sqrt{\left|Q_{0}\right|K(\phi)}\left(w_{m}+1\right).\label{eq:B equ 9}\end{align}
Since these equations are uncoupled in the leading order, the stability analysis
is performed for each equation separately. Integrating Eq.~(\ref{eq:B equ 9}),
we find the general solution\begin{equation}
B(\phi)=\exp\left[C_{0}-\frac{3\kappa\sqrt{\left|Q_{0}\right|}}{v_{0}}\left(w_{m}+1\right)\int^{\phi}\negmedspace\sqrt{K(\phi)}d\phi\right],\end{equation}
where $C_{0}$ is an integration constant. The general solution $B(\phi)$ will
tend to zero if and only if $\int\negmedspace\sqrt{K(\phi)}d\phi$ diverges
as $\phi\rightarrow\infty$ and $w_{m}>-1$. Let us temporarily denote \begin{equation}
\chi(\phi)\equiv\int^{\phi}\negmedspace\sqrt{K(\phi)}d\phi,\quad\chi\rightarrow\infty\:\textrm{as}\:\phi\rightarrow\infty.\label{eq:chi condition 9}\end{equation}
Then we rewrite the first equation as\begin{equation}
\frac{d}{d\chi}\left(-A\right)^{n-1}=-\frac{\left|Q_{0}\right|}{nv_{0}Q_{1}}\frac{K'(\phi)}{K^{3/2}}-\left(-A\right)^{n-1}\frac{3\kappa}{v_{0}}\sqrt{\left|Q_{0}\right|}.\end{equation}
(Note that we must have $A<0$ on solutions, due to the requirement of positivity
of $c_{s}^{2}$.) The general solution $A(\chi)$ can now be written explicitly,
but it suffices to observe that $A(\chi)$ will approach zero as $\chi\rightarrow\infty$
if and only if\begin{equation}
\lim_{\phi\rightarrow\infty}\frac{K'(\phi)}{K^{3/2}}=-2\lim_{\phi\rightarrow\infty}\frac{d}{d\phi}K^{-1/2}=0.\end{equation}
This condition is equivalent to\begin{equation}
\lim_{\phi\rightarrow\infty}\phi\sqrt{K(\phi)}=\infty.\label{eq:K prime condition 9}\end{equation}

Note that the condition~(\ref{eq:chi condition 9}) follows from that of Eq.~(\ref{eq:K prime condition 9}).
To verify this more formally, consider a function $K(\phi)$ such that $\int^{\infty}\negmedspace\sqrt{K(\phi)}d\phi<\infty$.
Then $K^{1/2}(\phi)$ necessarily decays faster than $\phi^{-1}$ at $\phi\rightarrow\infty$,
and so $K^{-1/2}$ grows faster than $\phi$ at $\phi\rightarrow\infty$. Such
$K(\phi)$ cannot satisfy Eq.~(\ref{eq:K prime condition 9}). Therefore it
is sufficient to impose only the condition~(\ref{eq:K prime condition 9}).
This condition is satisfied, for instance, by functions $K(\phi)\propto\phi^{s}$
with $s>-2$. Thus, we conclude that the solution with $R_{0}=0$ is asymptotically
stable under the condition~(\ref{eq:K prime condition 9}) and assumptions
$Q(v_{0})\neq0$, $Q'(v_{0})=0$.%
\footnote{This is case 8 in Sec.~\ref{sec:Viable-Lagrangians-for}.%
}

It remains to consider the case $R_{0}=0$, $Q(v_{0})=Q'(v_{0})=0$. In that
case, similarly to that discussed in Sec.~\ref{sub:Energy-density-2}, we may
use the expansions~(\ref{eq:expansion Q})--(\ref{eq:expansion c s}). It follows
that $w_{\phi}(v_{0})=0$. With the ansatz $v(\phi)=v_{0}-A(\phi)$, we find
that $A(\phi)<0$ on physically reasonable solutions. Then the leading terms
of Eq.~(\ref{eq:v equ clever}) are\begin{align}
\frac{d\left(-A\right)}{d\phi} & =-\frac{\left(-A\right)}{n-1}\frac{K'}{K}-\frac{3\kappa\sqrt{Q_{0}}}{\left(n-1\right)v_{0}}\sqrt{K(\phi)}\left(-A\right)^{\left(n+1\right)/2}.\label{eq:A equ 10}\end{align}
Since this equation is independent of $B$, it suffices to ensure that $A(\phi)\rightarrow0$
as $\phi\rightarrow\infty$ and subsequently consider the general solution for
$R(\phi)$. The general solution for $A(\phi)$ can be easily found by rewriting
Eq.~(\ref{eq:A equ 10}) as\begin{equation}
\frac{d}{d\phi}\left[\left(-A\right)^{-\left(n-1\right)/2}K^{-1/2}\right]=\frac{3\kappa\sqrt{Q_{0}}}{2v_{0}}.\end{equation}
We find\begin{equation}
\left(-A\right)^{\left(n-1\right)/2}=\frac{2v_{0}}{3\kappa\sqrt{Q_{0}}}\frac{1}{\phi-\phi_{0}}\frac{1}{\sqrt{K(\phi)}},\end{equation}
where $\phi_{0}$ is a constant of integration. It follows that $A(\phi)\rightarrow0$
as $\phi\rightarrow\infty$ if $K(\phi)$ is such that $\phi^{2}K(\phi)\rightarrow\infty$.
Under this assumption, we find that\begin{equation}
\varepsilon_{\phi}=K(\phi)\tilde{\varepsilon}_{\phi}(v)\propto\phi^{-2},\quad\phi\rightarrow\infty,\end{equation}
as it should according to Eq.~(\ref{eq:epsilon k asympt 44}). Now we analyze
the general solution for $R(\phi)$ Then the leading terms of Eq.~(\ref{eq:R EOM})
are\begin{equation}
\frac{dR}{d\ln\phi}=-R\frac{3\kappa\sqrt{E_{0}}}{v_{0}}\left(w_{m}+\frac{A}{nv_{0}}\right)=-2R\left(w_{m}+\frac{A}{nv_{0}}\right),\end{equation}
where we used Eq.~(\ref{eq:E0 cond 4}). If $w_{m}=0$, the right-hand side
above is always positive and (since $R$ is always positive) the general solution
for $R(\phi)$ cannot approach zero. If $w_{m}\neq0$, the general solution
for $R(\phi)$ is\begin{equation}
R(\phi)\propto\phi^{-2w_{m}}\quad\textrm{as}\,\phi\rightarrow\infty.\end{equation}
It follows that the general solution $R(\phi)\rightarrow0$ at $\phi\rightarrow\infty$
as long as $w_{m}>0$.  We conclude that an asymptotically stable solution
exists in case $Q(v_{0})=Q'(v_{0})=0$ if $w_{m}>0$ and $\phi^{2}K(\phi)\rightarrow\infty$
as $\phi\rightarrow\infty$. The admissible functions $K(\phi)$ are, for instance,
$K(\phi)\propto\phi^{s}$ with $s>-2$.%
\footnote{This is case 9 in Sec.~\ref{sec:Viable-Lagrangians-for}.%
}

\subsection{Slow motion ($v_{0}=0$), main case ($Q(0)\neq0$)\label{sub:Slow-motion-main-case}}

Previously we have been assuming that $v_{0}\neq0$. Now we turn to the case
$v_{0}=0$, which means that the velocity $\dot{\phi}\equiv v(\phi)$ of the
field $\phi$ approaches zero, albeit sufficiently slowly so that $\phi$ still
reaches arbitrarily large values at late times. We will now obtain the conditions
for the existence of asymptotically stable solutions with $v(\phi)\rightarrow0$
at $\phi\rightarrow\infty$.

The finiteness of the speed of sound at $v\rightarrow0$,\begin{equation}
\lim_{v\rightarrow0}c_{s}^{2}(v)=\lim_{v\rightarrow0}\frac{Q'(v)}{vQ^{\prime\prime}(v)}<\infty,\end{equation}
requires that $Q'(0)=0$. Since the important quantity $\tilde{\varepsilon}_{\phi}(v)=vQ'-Q$
approaches $-Q(0)$ at late times, it is useful to distinguish two possibilities:
$Q(0)\neq0$ and (less generically) $Q(0)=0$. In this section we consider the
generic case, $Q(0)\equiv-Q_{0}\neq0$. Positivity of the energy density requires
that $K(\phi)Q_{0}>0$, and we will choose $K(\phi)>0$ and $Q_{0}>0$.

Under these assumptions, we may expand the function $Q(v)$ near $v=0$ as\begin{equation}
Q(v)=-Q_{0}+Q_{1}v^{n}\left[1+O(v)\right],\label{eq:Q expansion 3}\end{equation}
where $n\geq2$ is the lowest order of the nonvanishing derivative of $Q(v)$
at $v=0$, and $Q_{1}>0$ because $Q(v)$ is a convex and monotonically growing
function of $v$. Other relevant quantities are then expanded as\begin{align}
\tilde{\varepsilon}_{\phi}(v) & =Q_{0}+\left(n-1\right)Q_{1}v^{n}\left[1+O(v)\right],\\
w_{\phi}(v) & =-1+\frac{nQ_{1}}{Q_{0}}v^{n}\left[1+O(v)\right],\label{eq:wk expansion 3}\\
c_{s}^{2}(v) & =\frac{1}{n-1}\left[1+O(v)\right].\label{eq:cs expansion 3}\end{align}
It follows that the only possible equation of state is $w_{\phi}(0)=-1$, indicating
a possible de Sitter tracker solution.

The equations of motion~(\ref{eq:v equ clever})--(\ref{eq:R EOM}) become
(neglecting terms of order $v$)\begin{align}
\frac{dv}{d\phi} & =-\frac{1}{n-1}\left[\frac{Q_{0}}{nQ_{1}v^{n-1}}\frac{K'}{K}+3\kappa\sqrt{\frac{K(\phi)Q_{0}}{1-R}}\right],\label{eq:v equ 11}\\
\frac{dR}{d\phi} & =-\frac{3\kappa}{v}R\sqrt{1-R}\sqrt{K(\phi)Q_{0}}\left(w_{m}+1\right).\label{eq:R equ 11}\end{align}
The first step is to investigate the possibility that $R(\phi)\rightarrow1$
at large $\phi$ (we will find that this possibility cannot be realized). We
note that for $w_{m}>-1$, the right-hand side of Eq.~(\ref{eq:R equ 11})
always remains negative. Thus, for $w_{m}>-1$ the general solution $R(\phi)$
cannot tend to 1 at $\phi\rightarrow\infty$, regardless of the behavior of
$K(\phi)$ and $v(\phi)$. In case $w_{m}<-1$, we need to do more work to establish
that there are no asymptotically stable solutions with $R_{0}=1$.

Substituting the ansatz $R(\phi)=1-B(\phi)$ into Eq.~(\ref{eq:R equ 11})
and assuming that $B\rightarrow0$, we obtain (omitting terms of order $v$
and $B$)\begin{equation}
\frac{d\sqrt{B}}{d\phi}=-\frac{3\kappa}{2v}\sqrt{K(\phi)Q_{0}}\left|1+w_{m}\right|.\label{eq:B equ 11x}\end{equation}
Changing the variable from $\phi$ to $\chi$ defined by \begin{equation}
\chi(\phi)\equiv\int^{\phi}\sqrt{K(\phi)}d\phi,\label{eq:chi to phi change 11}\end{equation}
we find\begin{equation}
\frac{d\sqrt{B}}{d\chi}=-\frac{3\kappa}{2v}\sqrt{Q_{0}}\left|1+w_{m}\right|.\label{eq:B equ 11}\end{equation}
There are now two possibilities: either the integral in Eq.~(\ref{eq:chi to phi change 11})
diverges at $\phi\rightarrow\infty$, or it converges. Accordingly, either $\chi\rightarrow\infty$
or $\chi\rightarrow\chi_{0}<\infty$ at $\phi\rightarrow\infty$. In case $\chi\rightarrow\infty$
at $\phi\rightarrow\infty$, we would have\begin{equation}
\lim_{\chi\rightarrow\infty}\frac{d\sqrt{B}}{d\chi}=0.\end{equation}
Since the right-hand side in Eq.~(\ref{eq:B equ 11}) tends to infinity at
$\phi\rightarrow\infty$, the case $\chi\rightarrow\infty$ is impossible. Thus,
the integral in Eq.~(\ref{eq:chi to phi change 11}) must converge at $\phi\rightarrow\infty$.
It follows that $K(\phi)\rightarrow0$ faster than $\phi^{-2}$ at $\phi\rightarrow\infty$,
and then we may express $K(\phi)$ through an auxiliary function $K_{0}(\phi)$
as\begin{equation}
K(\phi)=\phi^{-2}K_{0}(\phi),\quad\lim_{\phi\rightarrow\infty}K_{0}(\phi)=0.\end{equation}
Further, we rewrite Eq.~(\ref{eq:v equ 11}) as\begin{align}
\frac{dv}{d\ln\phi} & =\frac{1}{n-1}\left[\frac{Q_{0}}{nQ_{1}v^{n-1}}\left(2-\frac{d\ln K_{0}}{d\ln\phi}\right)\right.\nonumber \\
 & \quad\left.-3\kappa\sqrt{\frac{K_{0}(\phi)Q_{0}}{B}}\right].\label{eq:v equ 11 ln phi}\end{align}
By construction,\begin{equation}
\lim_{\phi\rightarrow\infty}\left(2-\frac{d\ln K_{0}}{d\ln\phi}\right)>2\label{eq:2 ln greater 2}\end{equation}
(the limit might even be positive infinite if $K_{0}$ tends to zero sufficiently
quickly). Hence, under the assumptions $v(\phi)\rightarrow0$ and $B(\phi)\rightarrow0$
we must have\begin{equation}
\lim_{\phi\rightarrow\infty}\frac{Q_{0}}{nQ_{1}v^{n-1}}\left(2-\frac{d\ln K_{0}}{d\ln\phi}\right)=+\infty.\end{equation}
It then follows by taking the limit $\phi\rightarrow\infty$ of Eq.~(\ref{eq:v equ 11 ln phi})
that the two terms in the brackets must cancel while both approach infinity.
Therefore, at large $\phi$ we must have the approximate relationship\begin{equation}
\frac{v^{n-1}(\phi)}{\sqrt{B(\phi)}}\approx\frac{\sqrt{Q_{0}}}{3\kappa nQ_{1}}\frac{1}{\sqrt{K_{0}(\phi)}}\left(2-\frac{d\ln K_{0}}{d\ln\phi}\right)\equiv M(\phi).\label{eq:M def 11}\end{equation}
Due to Eq.~(\ref{eq:2 ln greater 2}), the auxiliary function $M(\phi)$ defined
by Eq.~(\ref{eq:M def 11}) has the properties\begin{equation}
\lim_{\phi\rightarrow\infty}M(\phi)\sqrt{K_{0}(\phi)}>\frac{2\sqrt{Q_{0}}}{3\kappa nQ_{1}},\quad\lim_{\phi\rightarrow\infty}M(\phi)=+\infty.\label{eq:M property 2}\end{equation}
(The first limit may be positive infinite.) Using the function $M(\phi)$, we
may rewrite Eq.~(\ref{eq:v equ 11 ln phi}) as\begin{equation}
\frac{dv}{d\ln\phi}=\frac{3\kappa\sqrt{K_{0}(\phi)Q_{0}}}{n-1}\left[\frac{M}{v^{n-1}}-\frac{1}{\sqrt{B}}\right].\end{equation}
Expressing $\sqrt{B}$ through $v$ using Eq.~(\ref{eq:M def 11}) and substituting
the resulting expression for $\sqrt{B}$ into Eq.~(\ref{eq:B equ 11x}), we
find\begin{align}
\frac{d}{d\ln\phi}\left[\frac{v^{n-1}}{M}\right] & =-\frac{3\kappa}{2v}\sqrt{K_{0}(\phi)Q_{0}}\left|1+w_{m}\right|\nonumber \\
 & =\left(n-1\right)\frac{v^{n-2}}{M}\frac{dv}{d\ln\phi}-v^{n-1}M^{-2}\frac{dM}{d\ln\phi}.\end{align}
Rewriting the last equation as\begin{equation}
\frac{3\kappa}{2}M\sqrt{K_{0}Q_{0}}\left|1+w_{m}\right|=-\frac{n-1}{n}\frac{dv^{n}}{d\ln\phi}+v^{n}\frac{d\ln M}{d\ln\phi},\end{equation}
we note that the left-hand side tends to a positive limit (or to a positive
infinity) due to Eq.~(\ref{eq:M property 2}), while the term $dv^{n}/d\ln\phi$
tends to zero at $\phi\rightarrow\infty$ and can be neglected. Therefore, for
large $\phi$ we obtain\begin{equation}
v^{n}\approx\frac{3\kappa}{2}\sqrt{K_{0}(\phi)Q_{0}}\left|1+w_{m}\right|M^{2}\left[\frac{dM}{d\ln\phi}\right]^{-1}.\end{equation}
This relationship is sufficient for our purposes; we will now show that $v(\phi)$
cannot tend to zero at $\phi\rightarrow\infty$. If we assume that $v(\phi)\rightarrow0$,
we must have\begin{equation}
\lim_{\phi\rightarrow\infty}\frac{\sqrt{K_{0}(\phi)}}{\frac{d}{d\ln\phi}M^{-1}}=0.\end{equation}
Using Eq.~(\ref{eq:M def 11}), we transform this condition into\begin{equation}
\lim_{\phi\rightarrow\infty}\left[\frac{\frac{1}{2}\left(\ln K_{0}\right)_{,\ln\phi}}{2-\left(\ln K_{0}\right)_{,\ln\phi}}+\frac{d}{d\ln\phi}\left[2-\frac{d\ln K_{0}}{d\ln\phi}\right]^{-1}\right]=\infty.\label{eq:ln K cond 11}\end{equation}
It is now straightforward to show that the condition~(\ref{eq:ln K cond 11})
cannot be satisfied by a function $K_{0}(\phi)$ that tends to zero at $\phi\rightarrow\infty$.
Since $\left(\ln K_{0}\right)^{\prime}\leq0$ for all $\phi$, the function
$\left(\ln K_{0}\right)_{,\ln\phi}$ tends to a nonpositive constant or to a
negative infinity at $\phi\rightarrow\infty$. Hence, we obtain the bounds\begin{equation}
-1<\frac{\frac{1}{2}\left(\ln K_{0}\right)_{,\ln\phi}}{2-\left(\ln K_{0}\right)_{,\ln\phi}}<0,\quad0<\left[2-\frac{d\ln K_{0}}{d\ln\phi}\right]^{-1}<\frac{1}{2}.\end{equation}
The derivative of a bounded function cannot have an infinite limit. Therefore
the limit~(\ref{eq:ln K cond 11}) cannot be infinite. Since the condition~(\ref{eq:ln K cond 11})
cannot be satisfied, solutions with $v(\phi)\rightarrow0$ and $B(\phi)\rightarrow0$
do not exist under the present assumptions. 

Having shown that $R_{0}=1$ is impossible, we assume $R_{0}<1$ in the rest
of this section. Let us now consider the admissible behavior of $v(\phi)$ at
large $\phi$. It is convenient to change the independent variable from $\phi$
to $\chi$ defined by Eq.~(\ref{eq:chi to phi change 11}) and to rewrite Eq.~(\ref{eq:v equ 11})
as \begin{equation}
\frac{dv}{d\chi}=\frac{1}{n-1}\left[\frac{Q_{0}}{nQ_{1}v^{n-1}}\left(\frac{2}{\sqrt{K}}\right)_{,\phi}-3\kappa\sqrt{\frac{Q_{0}}{1-R}}\right].\label{eq:v equ 11a}\end{equation}
For an asymptotically stable solution, we need $v(\phi)\rightarrow0$ while
$v(\phi)>0$. Therefore, $dv/d\phi$ (and therefore also $dv/d\chi$) must remain
negative at large $\phi$. Let us examine the condition under which the right-hand
side of Eq.~(\ref{eq:v equ 11a}) might be negative at large $\phi$. 

We notice that the first term in the right-hand side of Eq.~(\ref{eq:v equ 11a})
contains a negative power of $v$ multiplied by a nonnegative function $d(K^{-1/2})/d\phi$
and a positive constant. This term will diverge to positive infinity as $v\rightarrow0$
unless $d(K^{-1/2})/d\phi$ tends to zero at large $\phi$. On the other hand,
the second term,\begin{equation}
-3\kappa\sqrt{\frac{Q_{0}}{1-R}},\end{equation}
tends to a negative constant at large $\phi$. Thus, $dv/d\chi$ may become
negative at large $\phi$ only when $d(K^{-1/2})/d\phi$ tends to zero at large
$\phi$. If $K(\phi)$ is such that $K'K^{-3/2}\rightarrow0$, then $\chi(\phi)\equiv\int^{\phi}\sqrt{K(\phi)}d\phi$
diverges at $\phi\rightarrow\infty$; this was already shown in the previous
section after Eq.~(\ref{eq:K prime condition 9}). Let us therefore continue
the analysis under the assumptions~(\ref{eq:K prime condition 9}) and $R_{0}<1$,
taking into account that $\chi\rightarrow\infty$ together with $\phi\rightarrow\infty$.

Rewriting Eq.~(\ref{eq:R equ 11}) as\begin{equation}
\frac{dR}{d\chi}=-\frac{3\kappa}{v}R\sqrt{1-R}\sqrt{Q_{0}}\left(w_{m}+1\right)\left[1+O(v)\right],\label{eq:R equ 11a}\end{equation}
and noting that $w_{m}+1\neq0$, we immediately see that $dR/d\chi\rightarrow0$
can be realized only if $w_{m}+1>0$ and either $R_{0}=0$ or $R_{0}=1$, where
$R_{0}\equiv\lim_{\phi\rightarrow\infty}R(\phi)$. Since we are assuming $R_{0}<1$,
the only admissible value is $R_{0}=0$. Therefore, we now look for solutions
of Eqs.~(\ref{eq:v equ 11a})--(\ref{eq:R equ 11a}) such that $v(\chi)\rightarrow0$
and $R(\chi)\rightarrow0$ as $\chi\rightarrow\infty$ (at the same time as
$\phi\rightarrow\infty$). 

Computing the limit of Eq.~(\ref{eq:v equ 11a}) as $\phi\rightarrow\infty$
and noting that $dv/d\chi\rightarrow0$ on asymptotically stable solutions,
we obtain the condition\begin{equation}
\lim_{\phi\rightarrow\infty}\frac{Q_{0}}{nQ_{1}}\frac{1}{v^{n-1}}\left(\frac{2}{\sqrt{K}}\right)_{,\phi}=3\kappa\sqrt{Q_{0}}.\end{equation}
The right-hand side above is a nonzero constant. Therefore it suffices to look
for solutions $v(\phi)$ of the form \begin{equation}
v^{n-1}(\phi)=\frac{\sqrt{Q_{0}}}{3\kappa nQ_{1}}\left(\frac{2}{\sqrt{K}}\right)_{,\phi}\left[1+A(\phi)\right],\label{eq:v ansatz 11}\end{equation}
where $A(\phi)$ is a new unknown function replacing $v(\phi)$. Solutions $v(\phi)\rightarrow0$
will be asymptotically stable if the general solution for $A(\phi)$ tends to
zero as $\phi\rightarrow\infty$. For brevity, we rewrite the ansatz~(\ref{eq:v ansatz 11}),
with the independent variable $\phi$ expressed through $\chi$, as\begin{equation}
v(\chi)=\left[\left(1+A(\chi)\right)W(\chi)\right]^{\frac{1}{n-1}},\label{eq:v ansatz 11a}\end{equation}
where $W(\chi)$ is a fixed function defined through\begin{equation}
\left.W(\chi)\right|_{\chi=\chi(\phi)}=\frac{\sqrt{Q_{0}}}{3\kappa nQ_{1}}\left(\frac{2}{\sqrt{K}}\right)_{,\phi}.\end{equation}
By assumption, we have $W(\chi)\rightarrow0$ as $\chi\rightarrow\infty$. Substituting
the ansatz~(\ref{eq:v ansatz 11a}) into Eqs.~(\ref{eq:v equ 11a})--(\ref{eq:R equ 11a}),
we obtain, to the leading order in $A$ and $R$,\begin{align}
\frac{dA}{d\chi} & =-\frac{3\kappa\sqrt{Q_{0}}\left(\frac{1}{2}R+A\right)+\left(n-1\right)\left(W^{\frac{1}{n-1}}\right)_{,\chi}}{W^{\frac{1}{n-1}}},\\
\frac{dR}{d\chi} & =-3\kappa\sqrt{Q_{0}}\left(w_{m}+1\right)RW^{-\frac{1}{n-1}}.\end{align}
Since the equation for $R$ does not contain $A$, the stability analysis can
be performed first for $R(\chi)$ and then for $A(\chi)$ assuming that $R(\chi)\rightarrow0$.
It is convenient to replace the independent variable $\chi$ temporarily by\begin{equation}
\psi(\chi)\equiv\int^{\chi}W^{-\frac{1}{n-1}}(\chi)d\chi.\end{equation}
Since $W(\chi)\rightarrow0$ as $\chi\rightarrow\infty$, the new variable $\psi$
grows to infinity together with $\chi$. The new equations for $A(\psi)$ and
$R(\psi)$ are\begin{align}
\frac{dA}{d\psi} & =-3\kappa\sqrt{Q_{0}}\left(\frac{1}{2}R+A\right)-\left(n-1\right)\left(W^{\frac{1}{n-1}}\right)_{,\chi},\\
\frac{dR}{d\psi} & =-3\kappa\sqrt{Q_{0}}\left(w_{m}+1\right)R.\end{align}
It is clear that the general solution for $R(\psi)$ tends to zero if $w_{m}>-1$.
The general solution for $A(\psi)$ is a sum of the general homogeneous solution
(which tends to zero) and an inhomogeneous solution. The inhomogeneous terms
are proportional to $R$ and $\left(W^{1/(n-1)}\right)_{,\chi}$, both of which
tend to zero at $\chi\rightarrow\infty$ ($\psi\rightarrow\infty$). Therefore
the general solution for $v(\phi)$ and $R(\phi)$ is asymptotically stable
under the current assumptions.%
\footnote{This is case 10 in Sec.~\ref{sec:Viable-Lagrangians-for}.%
}

\subsection{Slow motion ($v_{0}=0$), marginal cases ($Q(0)=0$)}

Let us now turn to the case $Q(0)=0$. In this case, we may expand the relevant
quantities near $v=0$ as follows,\begin{align}
Q(v) & =Q_{1}v^{n}\left[1+O(v)\right],\label{eq:Q expansion 3a}\\
\tilde{\varepsilon}_{\phi}(v) & =\left(n-1\right)Q_{1}v^{n}\left[1+O(v)\right],\label{eq:epsilon expansion 3a}\\
w_{\phi}(v) & =\frac{1}{n-1}\left[1+O(v)\right],\label{eq:wk expansion 3a}\\
c_{s}^{2}(v) & =\frac{1}{n-1}\left[1+O(v)\right],\label{eq:cs expansion 3a}\end{align}
where $n\geq2$ and $Q_{1}>0$. Using these expansions, we rewrite the equations
of motion~(\ref{eq:v equ clever})--(\ref{eq:R EOM}), in the leading order
in $v$, as\begin{align}
\frac{dv}{d\phi} & =-\frac{v}{n}\frac{K'}{K}-\frac{3\kappa\sqrt{Q_{1}}}{\sqrt{n-1}}\sqrt{\frac{K(\phi)}{1-R}}v^{\frac{n}{2}},\label{eq:v equ 12}\\
\frac{dR}{d\phi} & =-3\kappa R\sqrt{1-R}\sqrt{\left(n-1\right)Q_{1}K(\phi)}\frac{w_{m}-w_{\phi}(v)}{v^{1-n/2}}.\label{eq:R equ 12}\end{align}
The possible asymptotic values of equation of state parameter $w_{\phi}(0)$
are $1/(n-1)$ for $n\geq2$; in particular, we can have $w_{\phi}(0)=\frac{1}{3}$,
mimicking radiation, if $n=4$. When $w_{m}=1/(n-1)$, we may need to expand
the term $w_{m}-w_{\phi}(v)$ to a higher nonvanishing order in $v$. For instance,
assuming an expansion\begin{equation}
Q(v)\equiv Q_{1}v^{n}+Q_{2}v^{n+p}\left[1+O(v)\right],\label{eq:Q expansion v2}\end{equation}
where $n\geq2$ and $p\geq1$, we find\begin{equation}
w_{\phi}(v)=\frac{Q(v)}{vQ'(v)-Q}=\frac{1+O(v)}{n-1}\left[1-\frac{pQ_{2}v^{p}}{\left(n-1\right)Q_{1}}\right].\label{eq:wk expansion 3b}\end{equation}

Let us begin by considering the possible asymptotic value $R_{0}=1$ of $R(\phi)$
at $\phi\rightarrow\infty$; values $R_{0}<1$ will be considered subsequently.
In case $R_{0}=1$, we write the ansatz $R(\phi)=1-B(\phi)$ and transform Eq.~(\ref{eq:R equ 12})
into\begin{equation}
\frac{d\sqrt{B}}{d\phi}=\frac{3}{2}\kappa\sqrt{\left(n-1\right)Q_{1}K(\phi)}v^{\frac{n}{2}-1}\left(w_{m}-w_{\phi}(v)\right).\label{eq:B equ 12a}\end{equation}
The right-hand side of the equation above must be negative to allow $\sqrt{B(\phi)}\rightarrow0$
at $\phi\rightarrow\infty$. This cannot happen if $w_{m}-w_{\phi}(0)>0$. Thus,
the only possibility for the existence of stable solutions is $w_{\phi}(v)>w_{m}$
for $v>0$ (which does not exclude $w_{\phi}(0)=w_{m}$). Under the assumption
$w_{\phi}(0)>w_{m}$, Eqs.~(\ref{eq:v equ 12}) and (\ref{eq:B equ 12a}) can
be rewritten (again keeping only the leading-order terms) as\begin{align}
 & \frac{d}{d\phi}\ln\left(K^{1/n}v\right)=-\frac{3\kappa\sqrt{Q_{1}}}{\sqrt{n-1}}\sqrt{\frac{K(\phi)}{B(\phi)}}v^{\frac{n}{2}-1},\label{eq:v equ 12b}\\
 & \frac{d\ln B}{d\phi}=-\frac{3\kappa\sqrt{Q_{1}}}{\sqrt{n-1}}\left(1-\left(n-1\right)w_{m}\right)\sqrt{\frac{K(\phi)}{B(\phi)}}v^{\frac{n}{2}-1}.\label{eq:B equ 12b}\end{align}
 In case $w_{\phi}(0)=w_{m}$, we assume the expansion~(\ref{eq:Q expansion v2})
and use Eq.~(\ref{eq:wk expansion 3b}); then Eq.~(\ref{eq:B equ 12b}) is
replaced by\begin{equation}
\frac{d\ln B}{d\phi}=\frac{3\kappa pQ_{2}}{\left(n-1\right)^{3/2}\sqrt{Q_{1}}}\sqrt{\frac{K(\phi)}{B(\phi)}}v^{\frac{n}{2}-1+p}.\label{eq:B equ 12c}\end{equation}
As in the case $w_{\phi}(0)\neq w_{m}$, stable solutions are possible only
if the right-hand side of Eq.~(\ref{eq:B equ 12c}) is negative, i.e.~if $Q_{2}<0$. 

We now need to analyze the solutions of the systems~(\ref{eq:v equ 12b})--(\ref{eq:B equ 12b})
and (\ref{eq:v equ 12b}), (\ref{eq:B equ 12c}) by looking for such $K(\phi)$
that the general solutions $v(\phi)$ and $B(\phi)$ always tend to zero in
the two cases.

The general solution of Eqs.~(\ref{eq:v equ 12b})--(\ref{eq:B equ 12b}) can
be found by first noticing that\begin{equation}
\frac{d}{d\phi}\left[\ln\left(K^{1/n}v\right)-\frac{\ln B}{1-\left(n-1\right)w_{m}}\right]=0.\end{equation}
Hence we may express \begin{equation}
B(\phi)=C_{0}\left[K^{1/n}v\right]^{1-\left(n-1\right)w_{m}},\end{equation}
where $C_{0}>0$ is an integration constant. Then we substitute this $B(\phi)$
into Eq.~(\ref{eq:v equ 12b}) and obtain the following equation for the auxiliary
function $u\equiv K^{1/n}v$,\begin{equation}
\frac{du}{d\phi}=-F(\phi)u^{s},\label{eq:u F s equ}\end{equation}
where we defined the auxiliary constant $s$ and function $F(\phi)$ as\begin{align}
s & \equiv\frac{n-1}{2}\left(1+w_{m}\right),\label{eq:s def through w and n}\\
F(\phi) & \equiv3\kappa\sqrt{\frac{Q_{1}}{\left(n-1\right)C_{0}}}K^{1/n}(\phi)>0.\label{eq:F def through K}\end{align}
Since by assumption $w_{m}<\frac{1}{n-1}$, the possible values of $s$ are
$s<\frac{n}{2}$. We are now looking for functions $K(\phi)$ such that both
$v=u/F$ and $B\propto u^{n-2s}$ always tend to zero as $\phi\rightarrow\infty$;
in other words, we require \begin{equation}
\lim_{\phi\rightarrow\infty}\frac{u(\phi)}{F(\phi)}=0,\quad\lim_{\phi\rightarrow\infty}u(\phi)=0\end{equation}
for the general solution $u(\phi)$. The general solution for $u(\phi)$ can
be written as\begin{equation}
u(\phi)=\begin{cases}
\left[C_{1}+\left(s-1\right)\int^{\phi}F(\phi)d\phi\right]^{1/(1-s)}, & s\neq1,\\
\exp\left(C_{1}-\int^{\phi}F(\phi)d\phi\right), & s=1,\end{cases}\label{eq:u sol 12}\end{equation}
where $C_{1}$ is a constant of integration. If $s<1$, the power $1/(1-s)$
is positive and so the general solution $u(\phi)$ does not tend to zero. If
$s\geq1$, the general solution $u(\phi)$ tends to zero in case $\int^{\phi}F(\phi)d\phi$
diverges as $\phi\rightarrow\infty$, and does not tend to zero if $\int^{\phi}F(\phi)d\phi$
converges. Therefore, the only possibility for a stable solution is $s\geq1$
and $\int^{\phi}F(\phi)d\phi\rightarrow\infty$ as $\phi\rightarrow\infty$,
or equivalently \begin{equation}
w_{m}>-\frac{n-3}{n-1};\quad\lim_{\phi\rightarrow\infty}\int^{\phi}K^{1/n}(\phi)d\phi=\infty.\label{eq:conditions 12}\end{equation}
It remains to examine the condition $u(\phi)/F(\phi)\rightarrow0$ under these
assumptions. 

Since we already have $u\rightarrow0$, the condition $u/F\rightarrow0$ holds
if $F(\phi)$ approaches a nonzero constant or infinity as $\phi\rightarrow\infty$.
However, if \begin{equation}
\lim_{\phi\rightarrow\infty}F(\phi)=\lim_{\phi\rightarrow\infty}K^{1/n}(\phi)=0,\end{equation}
the condition $u/F\rightarrow0$ is a nontrivial additional constraint on the
function $K(\phi)$. This constraint can be expressed as a condition on $K(\phi)$
as follows. We find from Eq.~(\ref{eq:u sol 12}) that\begin{equation}
\frac{u}{F}\propto\begin{cases}
\left[F^{s-1}\int^{\phi}F(\phi)d\phi\right]^{-1/(s-1)}, & s>1,\\
\exp\left(-\ln F-\int^{\phi}F(\phi)d\phi\right), & s=1.\end{cases}\end{equation}
The condition $u/F\rightarrow0$ is then equivalent to\begin{eqnarray}
\lim_{\phi\rightarrow\infty}F^{s-1}\int^{\phi}F(\phi)d\phi=\infty, &  & s>1;\label{eq:cond F 12 a}\\
\lim_{\phi\rightarrow\infty}\left(\ln F+\int^{\phi}F(\phi)d\phi\right)=\infty, &  & s=1.\label{eq:cond F 12 b}\end{eqnarray}
 We note that the left-hand sides in Eqs.~(\ref{eq:cond F 12 a})--(\ref{eq:cond F 12 b})
depend monotonically on the growth of $F(\phi)$; more precisely, the terms
under the limits become larger when we choose a faster-growing or slower-decaying
function $F(\phi)$. Thus, it is clear that the conditions~(\ref{eq:cond F 12 a})--(\ref{eq:cond F 12 b})
will hold if $F(\phi)$ decays sufficiently slowly as $\phi\rightarrow\infty$
(or grows, but this case was already considered). With some choices of $F(\phi)=F_{0}(\phi)$,
the limits in Eqs.~(\ref{eq:cond F 12 a})--(\ref{eq:cond F 12 b}) will be
finite nonzero constants. We can easily determine such $F_{0}(\phi)$,\begin{equation}
F_{0}(\phi)\propto\phi^{-1/s},\quad s\geq1,\;\phi\rightarrow\infty.\end{equation}
Hence, the limits~(\ref{eq:cond F 12 a})--(\ref{eq:cond F 12 b}) will be
infinite when $F(\phi)$ decays slower than $\phi^{-1/s}$. The corresponding
condition for $K(\phi)$ can be written as\begin{equation}
\lim_{\phi\rightarrow\infty}\phi^{n/s}K(\phi)=\infty.\label{eq:K condition 12}\end{equation}
We can make this argument more rigorous by assuming the ansatz\begin{equation}
K(\phi)=\phi^{-n/s}K_{1}(\phi),\end{equation}
where $K_{1}(\phi)>0$ is an auxiliary function. Note that the function $F(\phi)$
is related to $K(\phi)$ by Eq.~(\ref{eq:F def through K}), which contains
an arbitrary integration constant $C_{0}>0$. Thus we may write\begin{equation}
F(\phi)=C_{1}\phi^{-1/s}K_{1}^{1/n}(\phi),\end{equation}
where $C_{1}>0$ is an arbitrary constant. If \begin{equation}
\lim_{\phi\rightarrow\infty}K_{1}(\phi)=\infty,\end{equation}
 it means that $K_{1}(\phi)$ is larger than any constant at sufficiently large
$\phi$. Then we obtain lower bounds (for arbitrary constant $C_{2}>0$) \begin{eqnarray}
\int^{\phi}F(\phi)d\phi>C_{2}\phi^{1-1/s}, &  & s>1;\\
\int^{\phi}F(\phi)d\phi>C_{2}\ln\phi, &  & s=1,\end{eqnarray}
and the conditions~(\ref{eq:cond F 12 a})--(\ref{eq:cond F 12 b}) hold. On
the other hand, if \begin{equation}
\lim_{\phi\rightarrow\infty}K_{1}(\phi)\equiv K_{1}^{(0)}<\infty,\label{eq:K1 bad case 12}\end{equation}
we find\begin{eqnarray}
\int^{\phi}F(\phi)d\phi\approx C_{2}\phi^{1-1/s}, &  & s>1;\\
\int^{\phi}F(\phi)d\phi\approx C_{2}\ln\phi, &  & s=1,\end{eqnarray}
where $C_{2}>0$ is an arbitrary constant. In that case, the conditions~(\ref{eq:cond F 12 a})--(\ref{eq:cond F 12 b})
cannot hold for arbitrary $C_{2}$. Therefore, the condition~(\ref{eq:K condition 12})
is necessary and sufficient for Eqs.~(\ref{eq:cond F 12 a})--(\ref{eq:cond F 12 b})
to hold.

We conclude that an asymptotically stable solution exists for $v_{0}=0$, $Q(0)=0$
with the expansion~(\ref{eq:Q expansion v2}), when $R_{0}=1$, $w_{m}<\frac{1}{n-1}$,
and the conditions~(\ref{eq:conditions 12}) and (\ref{eq:K condition 12})
hold. We note that for $n=2$ the condition $w_{m}<\frac{1}{n-1}$ contradicts
the first condition in Eq.~(\ref{eq:conditions 12}), so admissible solutions
exist only for $n>2$.%
\footnote{This is case 6 in Sec.~\ref{sec:Viable-Lagrangians-for}.%
}

The remaining case requires the analysis of Eqs.~(\ref{eq:v equ 12b})--(\ref{eq:B equ 12c}).
The general solution of these equations cannot be obtained in closed form; however,
we only need to analyze the asymptotic behavior at $\phi\rightarrow\infty$.
So we will estimate the relative magnitude of different terms in these equations.
Let us rewrite Eqs.~(\ref{eq:v equ 12b})--(\ref{eq:B equ 12c}) as\begin{align}
\frac{d\ln v}{d\phi} & =-\frac{1}{n}\frac{K'}{K}-\tilde{Q}_{1}\frac{K^{1/2}}{\sqrt{B}}v^{n/2-1},\label{eq:v equ 12c}\\
\frac{d\sqrt{B}}{d\phi} & =-\tilde{Q}_{2}K^{1/2}v^{n/2-1+p},\label{eq:B equ 12d}\end{align}
where the auxiliary positive constants\begin{equation}
\tilde{Q}_{1}\equiv\frac{3\kappa\sqrt{Q_{1}}}{\sqrt{n-1}},\quad\tilde{Q}_{2}\equiv-\frac{3\kappa pQ_{2}}{\left(n-1\right)^{3/2}\sqrt{Q_{1}}}\end{equation}
were introduced for brevity. (Positivity of these constants is clearly necessary
for the existence of asymptotically stable solutions.) Suppose that $v(\phi)$
and $B(\phi)$ are decaying solutions of Eqs.~(\ref{eq:v equ 12c})--(\ref{eq:B equ 12d}),
and let us compare the magnitude of the terms in the right-hand side of Eq.~(\ref{eq:v equ 12c})
in the limit $\phi\rightarrow\infty$. There are only three possibilities: the
first term dominates; the two terms have the same order; or the second term
dominates. In other words, the limit of the ratio of the second term to the
first,\begin{equation}
q\equiv\lim_{\phi\rightarrow\infty}\frac{\tilde{Q}_{1}K^{3/2}v^{n/2-1}}{K'\sqrt{B}},\end{equation}
must be either zero, or finite but nonzero, or infinite. The value of $q$ must
be the same for every decaying solution $\left\{ v(\phi),B(\phi)\right\} $
except perhaps for a discrete subset of solutions, which we may ignore for the
purposes of stability analysis. In each of the three cases, Eqs.~(\ref{eq:v equ 12c})--(\ref{eq:B equ 12d})
are simplified and become amenable to asymptotic analysis in the limit $\phi\rightarrow\infty$.
We will now consider these three possible values of $q$ in turn.

If $q=0$, we have at large $\phi$\begin{equation}
\frac{1}{n}\frac{K'}{K}\gg\tilde{Q}_{1}\frac{K^{1/2}}{\sqrt{B}}v^{n/2-1},\label{eq:12d cond 1}\end{equation}
and thus only the first term is left in Eq.~(\ref{eq:v equ 12c}),\begin{equation}
\frac{d\ln v}{d\phi}\approx-\frac{1}{n}\frac{K'}{K}\quad\Rightarrow\quad v\propto K^{-1/n}.\end{equation}
Decaying solutions have $v(\phi)\rightarrow0$; so a necessary condition is
$K^{-1/n}(\phi)\rightarrow0$ at $\phi\rightarrow\infty$. With this $v(\phi)$,
the condition~(\ref{eq:12d cond 1}) becomes\begin{equation}
\left(K^{-1/n}\right)^{\prime}\gg\frac{\tilde{Q}_{1}}{\sqrt{B}}.\end{equation}
However, this condition cannot be satisfied, since the left-hand side tends
to zero at large $\phi$ while the right-hand side tends to infinity because
$B\rightarrow0$. Thus, decaying solutions $v(\phi),B(\phi)$ are impossible
with $q=0$.

If $q\neq0$ and $\left|q\right|<\infty$, we consider $q$ as an unknown constant
that possibly depends on the solutions $v(\phi)$ and $B(\phi)$. At large $\phi$,
we have\begin{align}
q\frac{K'}{K} & \approx\tilde{Q}_{1}\frac{1}{\sqrt{B}}v^{n/2-1}K^{1/2},\label{eq:12d cond 2}\\
\frac{d\ln v}{d\phi} & \approx-\frac{K'}{K}\left(\frac{1}{n}+q\right).\end{align}
Therefore, \begin{equation}
v(\phi)\propto K^{-q-1/n}(\phi)\rightarrow0\label{eq:12d cond 2a}\end{equation}
since we need a decaying solution $v(\phi)$. With this $v(\phi)$, Eq.~(\ref{eq:12d cond 2})
yields\begin{equation}
\sqrt{B}\approx\frac{\tilde{Q}_{1}}{q}\frac{K^{1-\left(n/2-1\right)q+1/n}}{K'}.\end{equation}
For a decaying solution $B(\phi)\rightarrow0$, we thus must have\begin{equation}
\frac{1}{\sqrt{B}}\propto\frac{d}{d\phi}\left[K^{\left(n/2-1\right)q-1/n}\right]\rightarrow\infty\;\textrm{as}\;\phi\rightarrow\infty,\label{eq:12d cond 2bx}\end{equation}
 and in particular\begin{equation}
\left(\frac{n}{2}-1\right)q-\frac{1}{n}\neq0.\label{eq:12d cond 2b}\end{equation}
 Substituting the expressions for $v(\phi)$ and $\sqrt{B(\phi)}$ into Eq.~(\ref{eq:B equ 12d}),
we find\begin{align}
 & \frac{d\sqrt{B}}{d\phi}=-\tilde{Q}_{2}K^{\frac{1}{2}-\left(\frac{n}{2}-1+p\right)\left(q+\frac{1}{n}\right)}\nonumber \\
 & =\frac{d}{d\phi}\frac{\tilde{Q}_{1}}{q}\frac{K^{1-\left(\frac{n}{2}-1\right)q+\frac{1}{n}}}{K'}\nonumber \\
 & =\frac{\tilde{Q}_{1}}{q}K^{-\left(\frac{n}{2}-1\right)q+\frac{1}{n}}\left[1-\left(\frac{n}{2}-1\right)q+\frac{1}{n}-\frac{KK^{\prime\prime}}{K^{\prime2}}\right].\end{align}
This is now a closed equation for $K(\phi)$, which we may rewrite as\begin{align}
\left(\frac{K}{K'}\right)^{\prime} & =1-\frac{KK^{\prime\prime}}{K^{\prime2}}\nonumber \\
 & =\left[\left(\frac{n}{2}-1\right)q-\frac{1}{n}\right]-\frac{q\tilde{Q}_{2}}{\tilde{Q}_{1}}K^{-p\left(q+\frac{1}{n}\right)}.\end{align}
Due to the conditions~(\ref{eq:12d cond 2a}), (\ref{eq:12d cond 2b}), and
since $p>0$, the right-hand side above tends to a nonzero limit as $\phi\rightarrow\infty$,
namely\begin{equation}
\left(\frac{K}{K'}\right)^{\prime}\approx\left(\frac{n}{2}-1\right)q-\frac{1}{n}\equiv\frac{1}{\alpha}\neq0.\end{equation}
It follows that the only admissible form of the function $K(\phi)$ is\begin{equation}
K(\phi)\propto\phi^{\alpha},\quad\phi\rightarrow\infty.\end{equation}
However, this expression does not satisfy Eq.~(\ref{eq:12d cond 2bx}). Therefore,
asymptotically stable solutions are impossible.

In the last case, $q=\infty$, we may disregard the first term in Eq.~(\ref{eq:v equ 12c})
and obtain\begin{equation}
\frac{d\ln v}{d\phi}\approx-\tilde{Q}_{1}\frac{K^{1/2}}{\sqrt{B}}v^{n/2-1}.\end{equation}
Then we can rewrite Eq.~(\ref{eq:B equ 12d}) as\begin{align}
\frac{d\ln\sqrt{B}}{d\phi} & =-\tilde{Q}_{2}\frac{K^{1/2}}{\sqrt{B}}v^{n/2-1+p}=\frac{\tilde{Q}_{2}}{\tilde{Q}_{1}}v^{p}\frac{d\ln v}{d\phi}\nonumber \\
 & =\frac{\tilde{Q}_{2}}{p\tilde{Q}_{1}}\frac{d}{d\phi}v^{p}.\end{align}
This relationship between $B$ and $v$ can be integrated and yields\begin{equation}
\sqrt{B}=\exp\left[C_{1}+\frac{\tilde{Q}_{2}}{p\tilde{Q}_{1}}v^{p}\right],\end{equation}
where $C_{1}$ is a constant of integration. It follows that it is impossible
to find simultaneously decaying solutions $v(\phi)\rightarrow0$ and $B(\phi)\rightarrow0$
at $\phi\rightarrow\infty$. 

This concludes the consideration of the case $R_{0}=1$ and $w_{m}=\frac{1}{n-1}$,
in which case there are no asymptotically stable solutions.

We now turn to the analysis of the case $R_{0}<1$. We first note that the leading
terms of Eq.~(\ref{eq:v equ 12}) do not contain $R$ when $R\rightarrow R_{0}<1$.
Therefore the stability analysis can be performed for $R(\phi)$ and $v(\phi)$
separately. Using the ansatz $R(\phi)=R_{0}+B(\phi)$ and assuming a fixed solution
$v(\phi)$, we find that the right-hand side of Eq.~(\ref{eq:R equ 12}) is
independent of $B(\phi)$ if $0<R_{0}<1$. Therefore, general solutions $B(\phi)$
will not approach zero in case $R_{0}\neq0$. It remains to look for asymptotically
stable solutions $v(\phi)$ and $R(\phi)$ in case $R_{0}=0$.

In case $R_{0}=0$, we begin by analyzing the asymptotic behavior of $v(\phi)$.
Rewriting Eq.~(\ref{eq:v equ 12}) as\begin{equation}
\frac{du}{d\phi}=-\frac{3\kappa\sqrt{Q_{1}}}{\sqrt{n-1}}u^{\frac{n}{2}}K^{1/n},\quad u\equiv K^{1/n}v,\label{eq:v equ 13}\end{equation}
we find the approximate general solutions (valid only for large $\phi$)\begin{align}
u(\phi)=\exp\left[-3\kappa\sqrt{Q_{1}}\int_{\phi_{0}}^{\phi}K^{1/2}d\phi\right], & \quad\; n=2,\label{eq:u solution 13a}\\
u(\phi)=\left[\frac{3\kappa\sqrt{Q_{1}}}{\sqrt{n-1}}\int_{\phi_{0}}^{\phi}K^{1/n}d\phi\right]^{-\frac{2}{n-2}}, & \quad\; n>2,\label{eq:u solution 13b}\end{align}
where $\phi_{0}$ is an integration constant. The general solution $v=K^{-1/n}u$
should tend to zero as $\phi\rightarrow\infty$. We note that Eq.~(\ref{eq:v equ 13})
is similar to Eq.~(\ref{eq:u F s equ}) after the replacements\begin{equation}
F(\phi)\equiv\frac{3\kappa\sqrt{Q_{1}}}{\sqrt{n-1}}K^{1/n}(\phi),\quad s\equiv\frac{n}{2}.\label{eq:F def new}\end{equation}
Therefore, we may use the conclusion obtained after Eq.~(\ref{eq:u sol 12}),
with the caveat that $F(\phi)$ is presently related to $K(\phi)$ uniquely,
without an arbitrary proportionality factor. This was used to exclude the boundary
case~(\ref{eq:K1 bad case 12}), which is presently still allowed. Thus the
condition~(\ref{eq:K condition 12}) obtained above,\begin{equation}
\lim_{\phi\rightarrow\infty}\phi^{n/s}K(\phi)=\lim_{\phi\rightarrow\infty}\phi^{2}K(\phi)=\infty,\label{eq:K condition 13 sufficient}\end{equation}
 is now merely a \emph{sufficient} condition for the stability of the general
solution $v(\phi)$. In the boundary case,\begin{equation}
\lim_{\phi\rightarrow\infty}\phi^{2}K(\phi)\equiv K_{0},\quad0<K_{0}<\infty,\label{eq:boundary case 13}\end{equation}
we find\begin{align}
v(\phi)\propto\exp\left[\left(1-3\kappa\sqrt{Q_{1}K_{0}}\right)\ln\phi\right], & \quad\; n=2,\\
v(\phi)\approx\textrm{const}, & \quad\; n>2.\end{align}
Thus, the case~(\ref{eq:boundary case 13}) yields a stable solution for $v(\phi)$
when $n=2$ and $3\kappa\sqrt{Q_{1}K_{0}}>1$. (The possibility $3\kappa\sqrt{Q_{1}K_{0}}=1$
is unphysical because it requires an infinitely precise fine-tuning of the parameters
in the field Lagrangian.) Thus a sharp condition for the asymptotic stability
of $v(\phi)$ is\begin{align}
K(\phi)\geq\frac{1}{9\kappa^{2}Q_{1}}\phi^{-2}\quad\textrm{at }\phi\rightarrow\infty, & \quad n=2;\label{eq:K ans 12a}\\
\lim_{\phi\rightarrow\infty}\phi^{2}K(\phi)=\infty, & \quad n>2.\label{eq:K ans 12b}\end{align}
A weaker necessary condition is\begin{equation}
\int^{\infty}K^{1/n}(\phi)d\phi=\infty.\label{eq:K integral 13}\end{equation}

Let us now consider the stability of the general solution for $R(\phi)$. It
follows from Eq.~(\ref{eq:R equ 12}) that\begin{equation}
\frac{d\ln B}{d\phi}=-3\kappa\sqrt{\left(n-1\right)Q_{1}K(\phi)}v^{\frac{n}{2}-1}\left(w_{m}-w_{\phi}(v)\right).\end{equation}
This equation integrates to\begin{equation}
B(\phi)=B_{0}\exp\left[-\textrm{const}\cdot\int^{\phi}\left(w_{m}-w_{\phi}(v)\right)K^{\frac{1}{2}}v^{\frac{n}{2}-1}d\phi\right],\label{eq:B integral 13}\end{equation}
where $B_{0}$ is an integration constant. The general solution for $B(\phi)$
will tend to zero as long as the integral in Eq.~(\ref{eq:B integral 13})
diverges to a positive infinity at $\phi\rightarrow\infty$,\begin{equation}
\int^{\infty}\left(w_{m}-w_{\phi}(v)\right)K^{\frac{1}{2}}v^{\frac{n}{2}-1}d\phi=\infty.\label{eq:condition integral 13}\end{equation}
 A necessary condition for that is $w_{m}\geq\frac{1}{n-1}$. Precise constraints
on $K(\phi)$ for Eq.~(\ref{eq:condition integral 13}) can be obtained by
considering the cases $n=2$, $n\neq2$, $w_{m}=\frac{1}{n-1}$, and $w_{m}>\frac{1}{n-1}$
separately.

If $w_{m}>\frac{1}{n-1}$, the condition~(\ref{eq:condition integral 13})
holds when \begin{equation}
\int^{\infty}K^{\frac{1}{2}}v^{\frac{n}{2}-1}d\phi=\infty.\end{equation}
 If $n=2$, the above integral diverges due to the necessary condition~(\ref{eq:K integral 13}).
If $n>2$, we use the solution~(\ref{eq:u solution 13b}), where $u=K^{1/n}v$,
to obtain\begin{equation}
\int^{\infty}\negmedspace\negmedspace K^{\frac{1}{2}}v^{\frac{n}{2}-1}d\phi=\int^{\infty}\!\negmedspace d\phi\left[\int_{\phi_{0}}^{\phi}K^{1/n}(\phi_{1})d\phi_{1}\right]^{-1}\negmedspace K^{1/n}(\phi).\label{eq:K integral 13a}\end{equation}
Temporarily introducing the auxiliary function\begin{equation}
I(\phi)\equiv\int^{\phi}K^{1/n}(\phi)d\phi,\end{equation}
we note that $\lim_{\phi\rightarrow\infty}I(\phi)=\infty$ by Eq.~(\ref{eq:K integral 13}).
Therefore we express Eq.~(\ref{eq:K integral 13a}) through $I(\phi)$ and
obtain\begin{align}
\int^{\infty}K^{\frac{1}{2}}v^{\frac{n}{2}-1}d\phi & =\int^{\infty}\frac{I'(\phi)}{I(\phi)}d\phi\\
 & =\lim_{\phi\rightarrow\infty}\ln I(\phi)+\textrm{const}=\infty.\nonumber \end{align}
Therefore, the general solution $B(\phi)$ tends to zero with $w_{m}>\frac{1}{n-1}$
for any $n\geq2$ under the condition~(\ref{eq:K integral 13}).

When $w_{m}=\frac{1}{n-1}$, it follows from Eq.~(\ref{eq:Q expansion v2})
that the integrand in Eq.~(\ref{eq:B integral 13}) acquires an additional
factor proportional to $v^{p}$, where $p\geq1$. Therefore the general solution
for $B(\phi)$ will tend to zero only if \begin{equation}
\int^{\infty}K^{1/2}v^{n/2-1+p}d\phi=\infty,\label{eq:K integral 13b}\end{equation}
 where we need to substitute $v(\phi)=K^{-1/n}u(\phi)$ and $u(\phi)$ as given
by Eqs.~(\ref{eq:u solution 13a})--(\ref{eq:u solution 13b}).

Consider first the case $n=2$; we will now show that the condition~(\ref{eq:K integral 13b})
is incompatible with the earlier condition~(\ref{eq:K ans 12a}). Using the
solution~(\ref{eq:u solution 13a}), we can rewrite the condition~(\ref{eq:K integral 13b})
as\begin{equation}
\int^{\infty}\negmedspace d\phi\, K^{\left(1-p\right)/2}\exp\left[-3p\kappa\sqrt{Q_{1}}\int_{\phi_{0}}^{\phi}K^{1/2}d\phi_{1}\right]=\infty.\label{eq:K integral 13c}\end{equation}
By the condition~(\ref{eq:K ans 12a}), we have\begin{align}
K^{(1-p)/2} & <\textrm{const}\cdot\phi^{p-1},\\
\int_{\phi_{0}}^{\phi}K^{1/2}(\phi_{1})d\phi_{1} & >\sqrt{K_{0}}\ln\phi+\textrm{const}.\end{align}
Therefore the integral in Eq.~(\ref{eq:K integral 13c}) is bounded from above
by\begin{equation}
\textrm{const}\int^{\infty}\negmedspace d\phi\,\phi^{p-1-3p\kappa\sqrt{Q_{1}K_{0}}}=\textrm{const}\int^{\infty}\negmedspace d\phi\,\phi^{-1-\alpha}<\infty,\end{equation}
where we temporarily denoted\begin{equation}
\alpha\equiv\left(3\kappa\sqrt{Q_{1}K_{0}}-1\right)p>0,\end{equation}
and so the condition~(\ref{eq:K integral 13c}) cannot hold. 

It remains to consider the case $w_{m}=\frac{1}{n-1}$ and $n>2$. Using Eq.~(\ref{eq:u solution 13b}),
we rewrite the condition~(\ref{eq:K integral 13b}) as\begin{equation}
\int^{\infty}K^{\frac{1-p}{n}}(\phi)\left[\int_{\phi_{0}}^{\phi}K^{1/n}(\phi_{1})d\phi_{1}\right]^{-\frac{2p}{n-2}-1}d\phi=\infty.\label{eq:K integral 13d}\end{equation}
 According to Eq.~(\ref{eq:K ans 12b}), we must have\begin{equation}
K(\phi)>C_{0}\phi^{-2}\end{equation}
for any $C_{0}>0$ at large enough $\phi$; thus $K(\phi)$ should decay slower
than $\phi^{-2}$. However, it is straightforward to verify that a power-law
behavior\begin{equation}
K(\phi)\propto\phi^{-2+\delta},\quad\phi\rightarrow\infty,\;\delta>0,\end{equation}
yields a convergent integral in Eq.~(\ref{eq:K integral 13d}). Therefore,
the only possibility of having an asymptotically stable solution is to choose
$K(\phi)$ such that it decays slower than $\phi^{-2}$ but faster than $\phi^{-2+\delta}$
for any $\delta>0$. An example of an admissible choice of $K(\phi)$ is\begin{equation}
K(\phi)\propto\phi^{-2}\left(\ln\phi\right)^{\alpha},\quad\alpha>0.\label{eq:K 13e example}\end{equation}
With this $K(\phi)$, we obtain the following asymptotic estimate at large $\phi$,\begin{equation}
\int_{\phi_{0}}^{\phi}K^{1/n}(\phi_{1})d\phi_{1}\propto\textrm{const}\cdot\phi^{1-2/n}\left(\ln\phi\right)^{\alpha/n},\end{equation}
and so the integral~(\ref{eq:K integral 13d}) becomes, after some algebra,\begin{equation}
\textrm{const}\int^{\infty}\phi^{-1}\left(\ln\phi\right)^{-\alpha\frac{p}{n-2}}d\phi=\infty\quad\textrm{if }\alpha\frac{p}{n-2}\leq1.\end{equation}
Since the convergence of the integral in Eq.~(\ref{eq:K integral 13d}) monotonically
depends on the growth properties of the function $K(\phi)$, it is clear that
the condition~(\ref{eq:K integral 13d}) will also hold for functions $K(\phi)$
satisfying Eq.~(\ref{eq:K ans 12b}) but growing slower than those given in
Eq.~(\ref{eq:K 13e example}). However, the condition~(\ref{eq:K integral 13d})
may not hold for $K(\phi)$ growing faster than those in Eq.~(\ref{eq:K 13e example}). 

To investigate the admissible class of functions $K(\phi)$ more precisely,
let us use the ansatz\begin{equation}
K(\phi)=\phi^{-2}K_{0}(\phi),\end{equation}
where $K_{0}(\phi)$ is a function growing slower than any power of $\phi$.
Then we have an asymptotic estimate (for $n>2$)\begin{equation}
\int_{\phi_{0}}^{\phi}K^{1/n}(\phi_{1})d\phi_{1}\approx\textrm{const}\cdot\phi^{1-2/n}\left(K_{0}(\phi)\right)^{1/n},\end{equation}
and we can rewrite Eq.~(\ref{eq:K integral 13b}) as\begin{align}
 & \int^{\infty}K^{\frac{1-p}{n}}(\phi)\left[\int_{\phi_{0}}^{\phi}K^{1/n}(\phi_{1})d\phi_{1}\right]^{-\frac{2p}{n-2}-1}d\phi\nonumber \\
 & =\textrm{const}\cdot\int^{\infty}\phi^{-1}\left[K_{0}(\phi)\right]^{-\frac{p}{n-2}}d\phi=\infty.\label{eq:K1 condition 13}\end{align}
Substituting $K_{0}=\phi^{2}K$ into Eq.~(\ref{eq:K1 condition 13}), we find
that the conditions~(\ref{eq:K ans 12b}) and (\ref{eq:K integral 13d}) are
equivalent to\begin{equation}
\lim_{\phi\rightarrow\infty}\phi^{2}K(\phi)=\infty,\quad\int^{\infty}\negmedspace\phi^{-1-\frac{2p}{n-2}}\left[K\left(\phi\right)\right]^{-\frac{p}{n-2}}d\phi=\infty.\label{eq:K 13e answer}\end{equation}
 The condition~(\ref{eq:K ans 12b}) guarantees the stability of $v(\phi)$,
while Eq.~(\ref{eq:K integral 13d}) guarantees the stability of $B(\phi)$.
Therefore, Eq.~(\ref{eq:K 13e answer}) is a sharp (necessary and sufficient)
condition for the stability of the solution $\left\{ v,B\right\} $.

A sufficient (but not a necessary) condition for the divergence of the integral
in Eq.~(\ref{eq:K1 condition 13}) is\begin{equation}
\lim_{\phi\rightarrow\infty}\left(\ln\phi\right)^{-\frac{n-2}{p}}K_{0}(\phi)<\infty.\end{equation}
The corresponding sufficient condition for $K(\phi)$ is\begin{equation}
\lim_{\phi\rightarrow\infty}\phi^{2}K(\phi)=\infty,\quad\lim_{\phi\rightarrow\infty}\phi^{2}\left(\ln\phi\right)^{-\frac{n-2}{p}}K(\phi)<\infty.\label{eq:K 13f answer}\end{equation}
The sharp condition~(\ref{eq:K 13e answer}) cannot be restated in terms of
the asymptotic behavior of $K(\phi)$ at $\phi\rightarrow\infty$, but of course
one can check whether Eq.~(\ref{eq:K 13e answer}) holds for a given $K(\phi)$.
The condition~(\ref{eq:K 13e answer}) specifies a rather narrow class of functions;
however, we strive for generality and avoid prejudice regarding the possible
Lagrangians. 

In this section we have shown that asymptotically stable solutions exist with
$v_{0}=0$ and $Q(0)=0$ only in the following cases: (a) Asymptotic value $R_{0}=1$.
Expansion~(\ref{eq:Q expansion 3a}) holds with $Q_{1}>0$, determining the
value of $n$, which should be $n>2$; $-\frac{n-3}{n-1}<w_{m}<\frac{1}{n-1}$
according to Eq.~(\ref{eq:conditions 12}); and $K(\phi)$ satisfies Eq.~(\ref{eq:K condition 12}),
where $s$ is defined by Eq.~(\ref{eq:s def through w and n}).%
\footnote{This is case 6 in Sec.~\ref{sec:Viable-Lagrangians-for}.%
} There are no stable solutions when $w_{m}=\frac{1}{n-1}$ and expansion~(\ref{eq:Q expansion v2})
holds. (b) Asymptotic value $R_{0}=0$. Expansion~(\ref{eq:Q expansion 3a})
holds with $Q_{1}>0$, determining the value of $n\geq2$; either $n=2$, $w_{m}>1$,
and $K(\phi)$ satisfies Eq.~(\ref{eq:K ans 12a})%
\footnote{This is case 7 in Sec.~\ref{sec:Viable-Lagrangians-for}.%
} or Eq.~(\ref{eq:K ans 12b}),%
\footnote{This is case 11 in Sec.~\ref{sec:Viable-Lagrangians-for}.%
} or $n>2$, $w_{m}=\frac{1}{n-1}$, and $K(\phi)$ satisfies Eq.~(\ref{eq:K 13e answer}).%
\footnote{This is case 12 in Sec.~\ref{sec:Viable-Lagrangians-for}.%
}

\bibliographystyle{apsrev}
\bibliography{1}

\end{document}